\documentclass[1p]{elsarticle}

\usepackage{lineno,hyperref}

\journal{Cognitive Neurodynamics}

\usepackage{mathrsfs}
\usepackage{amssymb,amsthm}
\usepackage{amsmath}
\usepackage{mathtools}
\usepackage{caption}
\usepackage{subcaption}
\usepackage{xcolor,hyperref}
\usepackage{comment}
\usepackage[normalem]{ulem}

\theoremstyle{remark}
\newtheorem{remark}{\bf Remark}[section]

\def\w{{\omega}}
\def\X{{\textbf{X}}}
\def\x{{\mathbf{x}}}
\def\y{{\mathbf{y}}}
\def\Y{{\textbf{Y}}}
\def\Rn{{\mathbb{R}^{N}}}

\def\v{{\textbf{v}}}
\def\d{{\partial}}
\def\grad{{\nabla}}
\def\H{{\mathscr{H}}}
\def\J{{\mathscr{J}}}
\def\D{{\mathscr{D}}}
\def\V{{\mathscr{V}}}

\def\lam{{\lambda}}

\def\s{{\sigma}}
\def\Oo{{\mathcal O}}

\def\j{{\textbf{J}}}

\def\e{{\epsilon}}
\def\L{{\mathcal{L}}}
\def\b{{\boldsymbol\beta}}
\def\a{{\boldsymbol\alpha}}
\DeclareMathOperator*{\argmin}{argmin}

\newtheorem{thm}{Theorem}
\newtheorem{lem}[thm]{Lemma}
\newdefinition{rmk}{Remark}
\newproof{pf}{Proof}
\newproof{pot}{Proof of Theorem \ref{thm2}}

\begin{document}


\begin{frontmatter}

\title{Autonomous learning of nonlocal stochastic neuron dynamics}

\author[ucb]{Tyler E. Maltba\corref{cor1}\fnref{fn1}}

\author[ucb2]{Hongli Zhao\fnref{fn2}}

\author[stan]{Daniel M. Tartakovsky\corref{cor2}\fnref{fn3}}

\cortext[cor1]{First author}
\cortext[cor2]{Corresponding author}
\address[ucb]{Department of Statistics, UC Berkeley, Berkeley, CA 94720, USA.}  
\address[ucb2]{Department of Mathematics, UC Berkeley, Berkeley, CA 94720, USA.}  
\address[stan]{Department of Energy Resources Engineering, Stanford University, Stanford, CA 94305, USA.}

\fntext[fn1]{Supported by National Science Foundation Graduate Research Fellowship under Grant No. DGE 1752814}
\fntext[fn2]{Supported by UC Berkeley SURF L\&S Program}
\fntext[fn3]{Supported in part by Air Force Office of Scientific Research under award number FA9550-18-1-0474, and by a gift from TOTAL.}

\begin{abstract}
Neuronal dynamics is driven by externally imposed or internally generated random excitations/noise, and is often described by systems of random or stochastic ordinary differential equations. Such systems admit a distribution of solutions, which is (partially) characterized by the single-time joint probability density function (PDF) of system states. It can be used to calculate such information-theoretic quantities as the mutual information between the stochastic stimulus and various internal states of the neuron (e.g., membrane potential), as well as various spiking statistics. When random excitations are modeled as Gaussian white noise, the joint PDF of neuron states satisfies exactly a Fokker-Planck equation. However, most biologically plausible noise sources are correlated (colored). In this case, the resulting PDF equations require a closure approximation. We propose two methods for closing such equations: a modified nonlocal large-eddy-diffusivity closure and a data-driven closure relying on sparse regression to learn relevant features. The closures are tested for the stochastic non-spiking leaky integrate-and-fire and FitzHugh-Nagumo (FHN) neurons driven by sine-Wiener noise. Mutual information and total correlation between the random stimulus and the internal states of the neuron are calculated for the FHN neuron.
\end{abstract}


\end{frontmatter}

\section{Introduction}

Computational neuroscience often employs the theory of nonlinear dynamical systems to quantitatively understand complex neuronal processes~\cite{izhikevich}. A key aspect of the dynamical systems approach is the ability to discover relevant properties of brain function without accounting for every detail. It comprises coupled ordinary differential equations (ODEs), with the number of ODEs determining the dimensionality of a dynamical system. Examples of such neuron representations are the one-dimensional leaky integrate-and-fire (LIF) model~\cite{LIF}, two-dimensional FitzHugh-Nagumo (FHN)~\cite{FHN} and Morris-Lecar~\cite{ML} models, and four-dimensional Hodgkin-Huxley model~\cite{HH}. To account for apparent randomness of neuron behavior (e.g., due to synaptic, internal, and/or channel noise), such models include a stochastic noise/source term. To account for parametric uncertainty, the coefficients in such models might be treated as random variables. In both cases, such models admit distributions of solutions that are (partially) characterized by single-time joint probability density functions (PDFs) of system states. These PDFs can be either used directly to characterize the ensemble behavior of large populations of neurons~\cite{Gerstner} or post-processed to compute interspike-interval (ISI) distributions~\cite{Greenwood, lolov}, first exit time (FET) distributions~\cite{Alzubaidi, Tuckwell}, and the mutual information between different system states, or between a system state and a stochastic input~\cite{spikes}.

Synaptic noise reflects randomness in the release of neurotransmitters by synapses and random inputs from other neurons. Internal noise originates from the operations of ionic channels, and channel noise stems from the random switching of such channels.\footnote{Due to the large number of these channels, channel noise is often ignored by arguing that fluctuations average out~\cite{yamakou19}. However, channel noise in and of itself can change the behavior of neurons~\cite{white, Zhou_2020},  hence, it is important to study its effects on a neuron's dynamics.} Regardless of its genesis, a random source in ODEs (e.g., ionic current in the LIF and FHN models) is frequently treated as Gaussian white noise \cite{Greenwood}. While mathematically and computationally convenient, Gaussian white noise is biologically implausible since it is unbounded and uncorrelated. For that reason, sine-Wiener (SW) channel noise---a bounded and correlated (colored) non-Gaussian process---has been used to model signal transmission~\cite{Kang13}, resonance~\cite{Liang10}, and detection~\cite{Liang11} in nervous systems;  to induce spike death in excitable systems~\cite{guo}; and to detect a weak periodic signal in the FHN neuron~\cite{yao18}.  While our experiments focus solely on SW noise, the general methodology presented in Sections \ref{sec:pdf} and \ref{sec:numerics} is applicable for arbitrary colored noise processes that have Lebesgue measurable and almost surely bounded paths. Another common choice, among many others, is the Ornstein-Uhlenbeck process as seen in~\cite{Kang_2020, Guo_2011}.

One of the attractive features of Gaussian white noise is that the resulting ODE-based neuron model can be viewed not only as a Langevin system, but also as a formal stochastic differential equation (SDE) whose solution has a joint PDF that satisfies exactly the Fokker-Planck equation~\cite{risken1996,ksendal_2003}. This deterministic advection-diffusion partial differential equation (PDE), whose dimensionality is determined by the number of state variables/ODEs in the underlying stochastic model, is no longer exact in the presence of colored noise. A joint PDF of neuron states---or other quantities such as ISI or FET---can be estimated via (multilevel) Monte Carlo simulations \cite{Haskell, Rosenbaum, Alzubaidi, Giles}, but this procedure requires significant computational resources and sheds little light on the probabilistic dynamics of the neuron. More efficient methods of stochastic computation, e.g., polynomial chaos expansions and stochastic finite elements, do not provide a computational speed-up in this setting because colored noise is characterized by a short correlation length, leading to the so-called curse of dimensionality (see, e.g., the references in~\cite{Wang-2013-Probability}). The method of distributions~\cite{Tartakovsky-2015-Method} can be used to derive a PDF equation for the stochastic solution of an ODE driven by colored noise~\cite{Wang-2013-Probability, Maltba}. In the limit of the vanishing correlation length, i.e., when colored noise reduces to white noise, such a PDF equation becomes exact, reducing to the corresponding Fokker-Planck equation. Otherwise, it requires a closure approximation to be computable.

We propose two alternative approaches to construct such closures: theory-based nonlocal large-eddy-diffusivity (LED) closures~\cite{kraichnan-1987-eddy, Wang-2013-Probability, Maltba} and a data-driven closure relying on statistical inference, i.e., dictionary learning of relevant derivative features in PDF equations~\cite{bakarji}. To the best of our knowledge, both the method of distributions and statistical learning of PDF equations from a few Monte Carlo runs have not been used in computational neuroscience before. We investigate the relative performance of these strategies on two stochastic neuron models with varying degrees of complexity, the stochastic non-spiking leaky integrate-and-fire (NS-LIF) and FHN models. While standard  LIF models are by far the simplest biologically plausible neuron models, which are advantageous for simulating large coupled networks, they are limited in their ability to capture the basic topology of a neuron's excitability. The FHN model overcomes this limitation and captures key features of the Hodgkin-Huxley model. Our stochastic analogues of the  NS-LIF and FHN models account for SW channel noise.

In Section~\ref{sec:pdf}, we present the method of distributions in the context of Langevin-type systems driven by colored noise,
and describe alternative approaches to the construction of closures for resulting PDF equations. Our numerical strategies for solving these equations are discussed in Section~\ref{sec:numerics}. The method's performance and errors associated with the proposed closures are investigated for the stochastic NS-LIF and FHN models in Section~\ref{sec:applications}. This section also contains our strategy for computing mutual information between channel noise (or stochastic input current) and a neuron's membrane potential. Main conclusions derived from this study are summarized in Section~\ref{sec:conclusions}. 

\section{Method of Distributions} 
\label{sec:pdf}

Consider a system of random ordinary differential equations (RODEs),
 \begin{align} \label{ivp2}
  \frac{\text d\mathbf x(t,\omega)}{\text dt} = \mathbf v(\x(t,\omega), t; \boldsymbol\xi(t,\omega)), \qquad
  \mathbf x(0,\omega) = \mathbf x^{0}(\omega),
  \end{align}
that is to be solved on a time interval $(0, T_f]$ and holds for almost every $\omega \in \Omega$ in  an appropriate probability space $(\Omega, \mathscr{F}, \mathbb{P})$. The solution $\x(t,\w) = [ x_{1}(t,\w), \dots, x_{N}(t,\w)]^\top: (0,T_f]\times \Omega \to \mathbb R^N$ is an $\mathbb R^N$-valued stochastic process, 
and the initial state $\x^0$ is an $N$-dimensional random vector with the joint probability density function (PDF) $f_{\mathbf x_0}(\mathbf X_0) : \mathbb R^N \to \mathbb R^+$. The given deterministic function $\v = [ v_1, \dots, v_N]^\top: \mathbb R^N \to \mathbb R^N$, parameterized with  a set of $N_\text{par}$ random coefficients $\boldsymbol \xi(t,\omega) = [\xi_1(t,\omega),\dots,\xi_{N_\text{par}}(t,\omega)]^\top$, satisfies conditions guaranteeing the existence of a unique path-wise solution $\mathbf x(t,\omega)$ (see for instance \cite{strand,rode,kloeden}). The zero-mean random processes $\boldsymbol \xi(t,\omega)$ are characterized by a prescribed single-point joint PDF $f_{\boldsymbol\xi}(\boldsymbol\Xi;t)$ and a two-point covariance matrix $\mathbf C_{\boldsymbol\xi}$ with elements $C_{ij} \equiv \mathbb E [\xi_i(t_1,\omega) \xi_j(t_2,\omega)]$, where $i,j = 1,\dots,N_\text{par}$ and $t_1, t_2 \in (0,T_f]$. Here and below we use $\mathbb E[\cdot]$ and $\langle \cdot \rangle$ interchangeably to denote the ensemble mean. 

Let $\X = [X_1,\dots, X_N]^\top \in \Rn$ be a variable in the system's phase space. At any given time $t$, the state of the system is (partially) characterized by the joint probability $F_\mathbf{x}(\mathbf X;t) \triangleq \mathbb{P}[\x(t,\omega) \le \X]$. If this single-point joint cumulative distribution function $F_\mathbf{x}(\mathbf X;t)$ is differentiable with respect to all $X_i$, then the system is described by the corresponding single-point joint PDF $f_{\mathbf x}(\mathbf X;t)$.  Our goal is to derive a deterministic equation for the PDF $f_{\mathbf x}(\mathbf X;t)$ of the random solution $\mathbf x(t,\omega)$ to~\eqref{ivp2}.

The RODE solution $\mathbf x(t,\omega)$ has paths that are continuously differentiable with derivatives that are at most H{\"o}lder continuous. Hence, paths of the driving sample processes (e.g., $\boldsymbol\xi$) of~\eqref{ivp2} are at most H{\"o}lder continuous and need only be Lebesgue measurable and almost surely bounded so that~\eqref{ivp2} can be interpreted in the sense of Carath\'eodory. Therefore, the random coefficients $\xi_i(t,\omega)$ are correlated in time, i.e., $C_{ij} \not\sim \delta(t_1 - t_2)$ where $\delta(\cdot)$ is the Dirac delta function. If the coefficients $\xi_i(t,\omega)$ are uncorrelated in time (i.e., white noise), then~\eqref{ivp2} should be rewritten as a formal SDE~\cite{ksendal_2003} for which $f_{\mathbf x}(\mathbf X;t)$ satisfies exactly a Fokker-Plank (advection-diffusion) equation.

Systems of RODEs can always be rewritten as a formal multidimensional SDE by treating the coefficients $\xi_i(t,\omega)$ as states and adding their governing SDEs to the system of equations~\cite{rode}. Doing so would give a (possibly much)  higher-dimensional Fokker-Planck equation for the joint density of $[x_1, \hdots,x_N, \xi_1,\hdots,\xi_{N_{par}}]$, significantly increasing computational costs when compared to the method of distributions, which seeks only a governing equation for $f_{\mathbf x}(\mathbf X;t)$. Such an equation is henceforth referred to as a PDF equation; it is not unique and its derivation requires a closure approximation (e.g.,~\cite{Wang-2013-Probability} and the references therein).

In general, correlated inputs render PDF equations nonlocal in both phase space and time [ibid]. We focus on two interconnected challenges: numerical treatment of nonlocal PDF equations and statistical inference (i.e., learning) of the relevant closure terms in such equations from a few Monte Carlo realizations of~\eqref{ivp2}.

We use the implementation of the method of distribution that treats random noise $\boldsymbol{\xi}(t)$ exactly. This is in contrast to the implementations based on truncated expansions of random inputs, e.g., in a Karhunen-Lo\'eve series~\cite{Venturi-2012-A, venturi-2013-exact}. Such expansions are often not appropriate for colored noise $\boldsymbol{\xi}(t)$  due to the presence of short correlation lengths. Derivation of a PDF equation starts by defining an auxiliary functional~\cite{Tartakovsky-2015-Method}
 \begin{equation}
\Pi(\X, t; \w) \triangleq \delta(\x(t,\w) - \X) = \prod_{i=1}^{N} \delta(x_{i}(t,\w) - X_{i}). \label{rawpdf}
   \end{equation}
At any time $t$, its ensemble mean  over all realizations of $\x$ is the PDF $f_{\mathbf x}(\mathbf X;t)$:  
   \[
\mathbb E[ \Pi] \triangleq \int_{\Rn} \delta(\Y - \X) f_\mathbf{x}(\Y;t) \text d \Y = f_{\mathbf x}(\X;t).
 \]
We show in \ref{regularize} that $\Pi(\X, t; \w)$ obeys, in the sense of distributions, the linear conservation law
  \begin{align}\label{clpi}
 \frac{\d\Pi}{\d t} + \grad_{\X} \cdot [\v(\mathbf X,t;\omega)\Pi] = 0,  \qquad
 \Pi(\X,0) = \delta(\x^0(\omega) - \X).
 \end{align}
This equation describes, e.g., advection, in the phase space $\mathbf X$, of a passive scalar (with concentration $\Pi$) in the random velocity field $\mathbf v$. In that context, its stochastic averaging/homogenization has been the subject of multiple investigations~\cite{kraichnan-1987-eddy}. 

A typical homogenization procedure employs the Reynolds decomposition to represent the random inputs and outputs as the sums of their means and zero-mean fluctuations around these means. For example,  $\v = \langle \v  \rangle + \v'$ where 
\[
\langle \v  \rangle(\mathbf X,t) \triangleq \int_{\mathbb R^{N_\text{par}}} \v(\X,t; \boldsymbol\Xi) f_{\boldsymbol \xi}(\boldsymbol\Xi) \text d \boldsymbol \Xi
\quad\text{and}\quad 
\langle \v'  \rangle = \mathbf 0.
\]
Likewise, the Reynolds decomposition of $\Pi$ is $\Pi = f_{\mathbf x} + \Pi'$.  Substituting these decompositions into~\eqref{clpi} and taking the ensemble mean yields an unclosed PDF equation
 \begin{align}\label{pdfeq1} 
 \frac{\d  f_{\mathbf x} }{\d t} + \grad_{\X} \cdot (\langle\v\rangle  f_{\mathbf x} ) + \grad_{\X} \cdot \langle\v'\Pi'\rangle = 0, \qquad
 f_{\mathbf x} (\X; 0) =  f_{\mathbf x_0}(\X),
\end{align}
with vanishing free-space boundary conditions (though others may be considered). This equation is unclosed because it contains the unknown flux term $\langle \v'\Pi'\rangle$. Following~\cite{Maltba, bst}, we consider three alternative versions of the LED closure: local, semi-local, and nonlocal.
In all three formulations, the terminal problem (\ref{closuresection})
\begin{eqnarray}\label{term1}
 \frac{\text d\boldsymbol\chi}{\text ds} = \langle\mathbf  v(\boldsymbol\chi, s)\rangle,  \qquad
 \boldsymbol\chi(t) = \X, 
 \end{eqnarray}
and its associated flow  $\boldsymbol\chi(s) \equiv \boldsymbol\Phi(s; \X, t)$ play an essential role. Whenever possible, we omit $\w$ for notational convenience.

\subsection{Local LED closure}

The local closure is constructed by a direct application of the classical LED approach in fluid dynamics~\cite{kraichnan-1987-eddy}.  Under that approach, the cross covariance $\langle\v'\Pi'\rangle(\X,t)$ is split into advection and diffusion components,  
\begin{equation}\label{eq:LEDclass}
\langle\v'\Pi'\rangle \approx \boldsymbol{\mathscr{V}}(\X,t) f_\x(\X;t) - \boldsymbol{\mathscr{D}}(\X,t)\grad_{\X}\, f_\x(\X;t),  
\end{equation}
where the LED drift velocity $\boldsymbol{\mathscr{V}}$ and the LED diffusion tensor $\boldsymbol{\mathscr{D}}$ are given in~(\ref{leddrift}) and~(\ref{leddiffusion}), respectively.  Underpinning this closure is the assumption that the density $f_\x$ and its spatial derivatives vary slowly over a correlation-length interval $(t-\tau,t)$. Therefore,  this closure is expected to be valid for systems with short correlation lengths. 

\subsection{Nonlocal LED closure}

The nonlocal closure does not rely on short correlations to the extent of the local closure and is second-order accurate in $\sigma\tau$, i.e., the product of the noise intensity and correlation length (see \cite{kampen} and \ref{closuresection}). Under the nonlocal closure, the cross-covariance term is approximated by 
\begin{subequations} \label{nonlocalvp}
  \begin{align}
\langle\v'\Pi'\rangle \approx - \int_{0}^{t}   \J(s; \X,t)  \grad_{\boldsymbol\Phi} \cdot 
\Big(\langle\v'(\X,t)\v'^\top(\boldsymbol\Phi(s;\X, t),s) \rangle f_\x(\boldsymbol\Phi(s;\X, t);s)\Big)  \text ds, 
\end{align}
where 
\begin{eqnarray}
\J(s ; \X,t) =  \exp \left(- \int_s^t \grad_{\boldsymbol\chi} \cdot \langle\mathbf v(\boldsymbol\chi(r),r)\rangle  \text dr \right). \label{jacobian}
 \end{eqnarray}
 \end{subequations}

For a given dimension $N$,  the computational complexity of the nonlocal PDF method is dominated by  the work of solving the terminal problem~\eqref{term1} and evaluating the Jacobian determinant~(\ref{jacobian}). The latter is dramatically simplified if the flow divergence $\nabla_{\boldsymbol\chi} \cdot \langle\mathbf v\rangle$ is linear in $\boldsymbol\chi$, which occurs in such classical test problems as the nonlinear pendulum and the Duffing oscillator; analyses of such problems, e.g.,~\cite{cho}, are likely to give an overly optimistic assessment of PDF methods. The computational experiments reported in Section~\ref{sec:applications} are chosen to have nonlinear flow divergence $\nabla_{\boldsymbol\chi} \cdot \langle\mathbf v\rangle$.

\subsection{Semi-local LED closure}

The semi-local LED closures~\cite{Maltba, bst} aim to combine the simplicity and speed of the local closure with (most of) the accuracy of the nonlocal one. One such closure is derived in~\ref{app:semilocal},
\begin{align}
\langle\v'\Pi'\rangle \approx -  \int_{0}^{t} & \J(s ; \X,t) \exp((t-s)\j(\X,t))^\top \nabla_{\X} \cdot\Big( \langle\v'(\X,t)\v'^\top(\boldsymbol\Phi(s ;\X,t),s) \rangle \J^{-1}(s;\X,t) f_\x(\X;t) \Big) \text ds,   \label{ledmod1}
\end{align}
where $\mathbf J$ is the Jacobian of the mean-field velocity, $\mathbf J(\mathbf X,t) \triangleq \partial \langle \mathbf v(\mathbf X,t)\rangle/\partial \mathbf X$. Numerical evaluation of $\mathbf\Phi(s; \X, t)$  in $\J$ is often computationally expensive. To alleviate this cost, we approximate it by its terminal condition, $\mathbf\Phi(s; \X, t) \approx \X$, resulting in the approximate Jacobian determinant $\tilde{\J}$.

\begin{remark}\label{remark:Taylor}
As an alternative, one can approximate $\mathbf\Phi(s; \X, t)$ via a higher-order Taylor expansion around the terminal condition $(\X,t)$. Our numerical experiments in Section~\ref{sec:applications} reveal that this approximation gave no significant improvement in accuracy, while simultaneously presenting more numerical challenges than the terminal approximation. 
\end{remark}

Evaluation of the inverse approximate Jacobian determinant $\tilde{\J}^{-1}$ introduces additional challenges that would otherwise be absent if $\J^{-1}$ was known exactly, including, but not limited to, singularities and/or negative values of $\langle\v'\Pi'\rangle(\X,t)$. Many of these issues may be resolved by approximating $\tilde{\J}^{-1}$ with a nonnegative spline. Another option, which entirely circumvents such issues, is to use a simpler but slightly less accurate semi-local closure,
\begin{align} \label{ledmod2}
\langle\v'\Pi'\rangle \approx& -  \int_{0}^{t} \tilde{\J}(s ; \X,t) \exp((t-s)\j(\X,t))^\top  \nabla_{\X} \cdot\Big( \langle\v'(\X,t)\v'^\top(\boldsymbol\Phi(s ;\X,t),s) \rangle f_\x(\X;t) \Big)\text ds.
\end{align}
It is obtained by approximating $f_\x(\mathbf\Phi(s; \X, t);s) \approx f_\x(\X;t)$ in (\ref{nonlocalvp}), and  amounts to assuming the PDF $f_\x$ to be approximately constant over the interval $(t-\tau,t)$.

\subsection{Data-driven closures}
\label{sec:dd closures}

Depending on the model, the integrals in (\ref{leddrift}), (\ref{leddiffusion}), (\ref{ledmod1}), and (\ref{ledmod2}) may or not be analytically computable. If they cannot be solved analytically, their numerical evaluation can be computationally expensive, and prohibitively so in high dimensions.  As an alternative to the LED closures presented above, we explore empirical (data-informed) closures of~\eqref{pdfeq1}. The latter are constructed by representing the unknown term $\langle\v'\Pi'\rangle(\X,t)$ as a differential operator
\begin{align}\label{eq:ML}
\langle\v'\Pi'\rangle \approx \beta_0(\X,t) f_\x(\X;t) + \mathbf D(\X,t) \nabla_\X f_\x(\X;t)+ \text{higher-order derivatives},
\end{align}
where $\mathbf D$ is a second-order semi-positive definite tensor with elements $D_{ij}$ ($i,j = 1\dots,N$). The $M$ real-valued coefficients $\b = \{\beta_0,D_{11},\dots,D_{NN},\dots\}$ are learned from post-processed Monte Carlo realizations of~\eqref{ivp2} by solving the following optimization problem.

For simplicity, we consider a square spatial domain $\mathcal{D}\subset \Rn$ and discretize with a uniform square mesh $X_n^k = X_n^0 +k\Delta X$ for  $k \in\{1,K\}$ in each spatial dimension $n$. For the temporal domain $(0,T_f]$, we also take the uniform grid $t^l = l\Delta t$ for $l\in\{1,L\}$. Let $\hat{f}_\x \in \mathbb{R}^{K\times ...\times K \times L}$ with entries $\hat{f}_\x^{k_1...k_Nl}\equiv \hat{f}_\x(X_1^{k_1},...,X_N^{k_N};t^l)$ denote a joint PDF computed via post-processing of Monte Carlo realizations of~\eqref{ivp2} with a kernel density estimator. We emphasize that $\hat{f}_\x$ is not the converged Monte Carlo solution, $f_\text{MC}$, of~\eqref{ivp2}. Instead, $\hat{f}_\x$ is computed using only a small fraction of the number of trials needed to compute the converged solution $f_\text{MC}$ with a prescribed accuracy. Furthermore, we denote by $\hat{\mathcal L}$ a suitable numerical discretization (e.g., finite difference, finite volume, total variation regularized differentiation, etc.) of the differential operator $\mathcal L \equiv \grad_{\X} \cdot (\langle\v\rangle  f_{\mathbf x}  + \langle\v'\Pi'\rangle )$ in~\eqref{pdfeq1}, with $\langle\v'\Pi'\rangle$ given by~\eqref{eq:ML}. Finally, we introduce the discretized residual
\[\mathcal{R}_{k_1...k_Nl}(\b) \triangleq \frac{\partial \hat{f}_\x^{k_1...k_Nl}}{\partial t} + \hat{\L}(\hat{f}_\x^{k_1...k_Nl};\b);\]
an example construction is discussed in detail in Section \ref{sec:numerics}. The optimal variable coefficient vector $\b$ is one that satisfies 
 \begin{eqnarray}
 \hat{\b}(\X,t) = \argmin_{\b(\X,t)} \left\{\frac{1}{K^NL} \sum_{k_1=1}^K \hdots \sum_{k_N=1}^K\sum_{l=1}^L ||\mathcal{R}_{k_1...k_Nl}(\b)||_2^2+\lam||\b||_1^2\right\}, \label{min1}
\end{eqnarray}
where $||\cdot||_2$ is the $L_2$ norm, $||\cdot||_1$ is the $L_1$ norm penalty ensuring sparsification of features, and $\lam$ is a hyper-parameter iteratively chosen through a geometric sequence~\cite{brunton, schaeffer, bakarji}. 

In order to generalize the optimization problem in (\ref{min1}), one must balance a bias-variance tradeoff when designing a hypothesis class, i.e., choosing the order of derivatives in~\eqref{eq:ML}. A large hypothesis class, represented above by $M$ variable coefficients $\b$, minimizes bias at the cost of high variance.  Such a class is more likely to fit an appropriate operator to the data in $\hat{f}_\x$; however, at the cost of possibly making the equations too difficult to manipulate analytically. Choosing too simple of a hypothesis class may prevent generalization by automatically removing hypotheses with high variance, i.e., filtering out noise and outliers in the data. By choosing to include both variable coefficients (as opposed to constants) and the $L_1$ penalty in (\ref{min1}), we aim to balance the bias-variance tradeoff by respectively increasing the power of the model, and by introducing sparsification, which leads to more interpretable equations.

The hypothesis class is further constrained by the fact that~\eqref{pdfeq1} is an equation for the PDF of (\ref{ivp2}), which means that it must take the form of a master equation. As a consequence of Pawula's theorem~\cite{risken1996,kampen}, the (finite) Kramers-Moyal (KM) expansion of the PDF master equation does not contain any partial derivatives of order higher than two; otherwise, it could yield a PDF that either is negative or may not integrate to one. Hence, the order of derivatives in~\eqref{eq:ML} is restricted to one.


\section{Numerics}
\label{sec:numerics}

For any of the above closures, we solve (\ref{pdfeq1}) through operator splitting by successively considering the two equations
\begin{align}
\frac{\partial f_\x}{\partial t} + \nabla_\X \cdot \langle\mathbf{v}\rangle f_\x = \;& 0, \label{split1} \\
\frac{\partial f_\x}{\partial t} + \nabla_\X\cdot \langle \mathbf v'\Pi'\rangle = \;& 0. \label{split2}
 \end{align}
The conservative advection equation (\ref{split1}) is solved through the \texttt{Clawpack} package~\cite{clawpack} using a Lax-Wendroff discretization and MC limiter. In the $N=1$ dimensional setting,~\eqref{split2} is solved via a Crank-Nicolson time discretization. For dimensions higher than one, the modified Craig-Sneyd scheme presented in \cite{Hout13} is used with $\theta = 1/2$. This stable alternating direction implicit scheme treats mixed derivatives explicitly while maintaining second-order temporal accuracy. In all cases,~\eqref{split2} is spatially discretized with second-order central differencing. \texttt{Clawpack} provides two options for operator splitting, namely Godunov or Strang splitting. However, Strang splitting is not currently available for adaptive mesh refinement. For the experiment in Section \ref{LIF_sec}, the classic version (uniform spatial mesh) of \texttt{Clawpack} is used with Strang splitting since adaptive mesh refinement is not necessary. For the experiment in Section \ref{sec:FHN}, we use the adaptive version of \texttt{Clawpack} (AMRClaw) with Godunov splitting, which significantly reduces computational costs. Although Godunov splitting is only formally first-order accurate, a convergence study revealed that near second-order temporal accuracy is maintained, i.e., its temporal truncation errors range between $\Oo(\Delta t^{1.9})$ and $\Oo(\Delta t^2)$.

The computational domain is taken as $[-L,L]^N$ where $L>0$ is the smallest number large enough to ensure that the domain truncation with homogeneous boundary conditions generates only negligible errors, by which we mean that if the size of each dimension of the domain is increased 10 fold, the error improves only by 0.01 percent.\footnote{In Section~\ref{LIF_sec}, the solution to~\eqref{ivp2} is positive, so the domain is reduced to $(0,L]$ with a homogeneous boundary condition at $X=L$ and a reflective (i.e., zero-flux) boundary condition at $X=0$.} For a fixed time $t$, the error is defined in terms of the Kullback-Leibler (KL) divergence~\cite{Mackay}, 
\[
D_\text{KL}(f_\text{MC}||f_\x)(t) \triangleq \int_{\mathbb{R}^N} f_\text{MC}(\X;t) \ln\left(\frac{f_\text{MC}(\X;t)}{f_\x(\X;t)}\right) \text d\X,
\]
It quantifies the amount of information lost when $f_\x$ is used to approximate the high-resolution Monte Carlo solution $f_\text{MC}$, which is treated as a yardstick.

Each Monte Carlo realization is achieved by solving~\eqref{ivp2} with an adaptive fourth-order Runge-Kutta (RK4) scheme, treating the discretized stochastic fluctuation $\v'$ as a variable coefficient. It is well known that standard Runge-Kutta schemes can be implemented for RODEs; however, they typically do not maintain their traditional orders of accuracy (see for instance \cite{rode,kloeden} and the references therein). A convergence study revealed that the standard RK4 scheme maintained third-order accuracy for the experiments in Section \ref{sec:applications}. For each time step in the domain, a large number of Monte Carlo realizations are then post-processed with a Gaussian kernel density estimator to obtain the yardstick PDF $f_\text{MC}$ on the discretized spatial domain. The exact number of realizations used to compute the yardstick solution is problem dependent. It is determined by a convergence study and is explicitly given for each experiment in Section \ref{sec:applications}. The bandwidth of the kernel density estimator is chosen by Silverman's rule. In addition to $f_\text{MC}$, we also compute the data array $\hat{f}_\x$ first defined in Section~\ref{sec:dd closures}. The number of realizations used to compute $\hat{f}_\x$ typically ranges from $5\%$ to $20\%$ of those used for $f_\text{MC}$.

\begin{remark}
Other natural choices for measuring error of PDF solutions are the Wasserstein and total variation distances. Relative $L^p$ norms are typically poor choices for measuring error of PDFs. They often underrepresent the accuracy of closures, sometimes increasing the error by one order of magnitude even if many statistical measures such as mean, variance, skewness, kurtosis, and modality match.  
\end{remark}

\begin{remark}
Asymptotic ``plug-in'' estimators of bandwidth, such as Scott's or Silverman's rule, are computationally inexpensive. Yet, if the underlying distribution is heavy-tailed or multimodal, computing the yardstick PDF $f_\text{MC}$ and the data array $\hat{f}_\x$ may require an unusually large number of realizations. More robust approaches to bandwidth selection include various forms of cross-validation (CV) (e.g., k-fold CV~\cite{Hastie_2009}, one-sided CV~\cite{savchuk1,savchuk2}, etc.) and total variation regularization~\cite{Koenker_2007}. The latter allows for variable bandwidths, but has to solve a bandwidth optimization problem at each time step. 
\end{remark}

\subsection{LED closures}
\label{sec:numerics_led}
The local and semi-local formulations of the LED closures have the same computational complexity, while the nonlocal formulation is considerably more expensive~\cite{Maltba}. Evaluation of the cross-covariance term~\eqref{nonlocalvp} in the latter involves a numerical approximation of the integral, which requires the history of the solution up to the current time step. 
In the results presented below, the proposed local and semi-local approaches are four orders of magnitude faster than the standard Monte Carlo approach.

\subsection{Data-driven closures}
\label{sec:numerics_dd}

Numerical solutions of~\eqref{split1} and~\eqref{split2} with semi-local or nonlocal LED closures~\eqref{nonlocalvp}--\eqref{ledmod2} may become highly nontrivial, e.g., when $N>3$ and $\v'$ is spatially dependent. Our numerical experiments reported in Section \ref{sec:applications} reveal the total computational costs of the data-driven closures to be higher than that of the local and semi-local LED closures. Yet, the data-driven closures are still considerably cheaper than the standard Monte Carlo approach and, in contrast to the LED closures, are relatively easy to implement numerically.  This makes the data-driven closures an attractive alternative. Moreover, if the optimization problem~\eqref{min1} is properly constrained and not naively implemented, the bulk computational costs of the data-driven approach lies in generating the (relatively few) Monte Carlo realizations needed to assemble $\hat{f}_\x$.
The remainder of this section is devoted to solving (\ref{min1}).

\begin{remark}
A naive strategy for solving the optimization problem~\eqref{min1}, which involves the discretized variable coefficients  $\b(X_1^{k_1},...,X_N^{k_N},t^l)$ at each spatiotemporal grid point $(X_1^{k_1},...,X_N^{k_N},t^l)$, is prohibitively expensive without proper computational resources, particularly memory allocation. The experiment in Section~\ref{sec:FHN}, which has a relatively low-dimensional ($N=2$) phase space, illustrates the high-dimensional nature of this optimization problem. An accurate solution of~\eqref{split1} would require a spatial grid size of at most $\Delta X_1 = \Delta X_2 = 0.01$ at the highest level of mesh refinement. Although this is handled with an adaptive mesh when solving~\eqref{split1} and~\eqref{split2}, a uniform mesh is used for the optimization problem~\eqref{min1}. For a square domain, a uniform mesh of size $\Delta X_1 = \Delta X_2 = 0.01$ consisting of $1000$ elements in each direction would result in $10^6$ spatial grid cells. Given that the time step $\Delta t$ must satisfy the hyperbolic CFL condition, the total number of spatiotemporal grid points is on the order of $10^9$ for each of the $M$ coefficients $\b$. The resulting dimension of the optimization problem is $\mathcal{O}(MK^NL) = M10^9$, well beyond the limitations of a personal computer.
\end{remark}

To reduce the dimensionality of the optimization problem~\eqref{min1}, we use a polynomial basis expansion to represent the decision variables $\b(\X,t)$. In this setting, it is natural to directly reference the elements of $\b(\X,t)$ in terms of $\beta_0$ and $\mathbf{D}$ from~\eqref{eq:ML}. For a polynomial (e.g., Chebyshev, Legendre, etc.) basis $P_{q}(\cdot)$,
\begin{subequations}\label{cheb}
\begin{align}
\beta_0(\X,t) &\triangleq \sum_{q_1=1}^{Q_1} \hdots \sum_{q_N=1}^{Q_N} \sum_{r=1}^{R} \alpha_0^{q_1\hdots q_Nr} P_{q_1}(X_1)\hdots P_{q_N}(X_N)P_{r}(t), \\
D_{ij}(\X,t) &\triangleq \sum_{q_1=1}^{Q_1} \hdots \sum_{q_N=1}^{Q_N} \sum_{r=1}^{R} \alpha_{ij}^{q_1\hdots q_Nr} P_{q_1}(X_1)\hdots P_{q_N}(X_N)P_{r}(t).
\end{align}
\end{subequations}
The optimization problem~\eqref{min1} is now solved over the expansion coefficients $\alpha_0^{q_1\hdots q_Nr}$ and $\alpha_{ij}^{q_1\hdots q_Nr} \in \mathbb{R}$ for $i,j \in\{1,N\}$. This step reduces the optimization dimension from $\mathcal{O}(MK^NL)$ to $\mathcal{O}(MQ_1\hdots Q_NR)$, where each $Q_1, ..., Q_N,R < 10$. For the experiments in Section \ref{sec:applications} , the optimization dimension is reduced from $M10^9$ to approximately $M10^1$.

To formulate the regression problem, we take the divergence of~\eqref{eq:ML}. Ignoring higher-order derivatives yields
\begin{eqnarray} \label{eq:dict}
\grad \cdot\langle\v'\Pi'\rangle \approx \sum_{i=1}^N \d_{X_i}(\beta_0f_\x)  + \sum_{i,j=1}^N \d_{X_i}\left[D_{ij}(\d_{X_j}f_\x)\right].
\end{eqnarray}
We let $\mathcal{P}_{q_1\hdots q_Nr}^{k_1\hdots k_Nl} \triangleq   P_{q_1}(X_1^{k_1})\hdots P_{q_N}(X_N^{k_N})P_{r}(t^l)$ denote the discretized polynomial bases. We also build a target vector $\mathbf{V} \in \mathbb{R}^{K^NL}$ in~\eqref{split1} that contains the advection terms for each spatiotemporal grid point approximated with a Lax-Wendroff discretization and MC limiter. Its elements, denoted by the mulit-index $k_1\hdots k_Nl$, are given by
\[ V^{k_1\hdots k_Nl} \triangleq 
\left( \partial_t \hat{f}_\x + \grad_\X \cdot \langle \v\rangle\hat{f}_\x \right)(X_1^{k_1},\hdots,X_N^{k_N},t^l).
\]
After substituting in the discretized version of~\eqref{cheb} into~\eqref{eq:dict} and changing the order of summations, for a fixed spatiotemporal grid node with index $k_1\hdots k_Nl$, the discretized residual takes the form
\begin{eqnarray}
\mathcal{R}_{k_1...k_Nl} \triangleq V^{k_1...k_Nl} &+& \sum_{q_1\hdots q_Nr}\sum_i\d_{X_i}\left[\mathcal{P}_{q_1\hdots q_Nr}^{k_1\hdots k_Nl}\hat{f}_\x^{k_1\hdots k_Nl}\right]\alpha_0^{q_1\hdots q_Nr} \notag \\
&+& \sum_{i,j} \sum_{q_1\hdots q_Nr} \d_{X_i}\left[\mathcal{P}_{q_1\hdots q_Nr}^{k_1\hdots k_Nl}\left(\d_{X_j}\hat{f}_\x^{k_1\hdots k_Nl}\right)\right]\alpha_{ij}^{q_1\hdots q_Nr}.
\end{eqnarray}
Rewriting in matrix/vector form gives
\begin{eqnarray}\label{eq:res2}
\boldsymbol{\mathcal{R}} = \mathbf{V} + \mathbf{F}_0\boldsymbol{\alpha}_0 + \sum_{i,j} \mathbf{F}_{ij}\boldsymbol{\alpha}_{ij},
\end{eqnarray}
where the elements of each matrix are given by
\begin{eqnarray}
\mathbf{F}_0(k_1\hdots k_Nl, q_1\hdots q_Nr) &\triangleq& \sum_i\d_{X_i}\left[\mathcal{P}_{q_1\hdots q_Nr}^{k_1\hdots k_Nl}\hat{f}_\x^{k_1\hdots k_Nl}\right], \label{F0}\\
\mathbf{F}_{ij}(k_1\hdots k_Nl, q_1\hdots q_Nr) &\triangleq& \d_{X_i}\left[\mathcal{P}_{q_1\hdots q_Nr}^{k_1\hdots k_Nl}\left(\d_{X_j}\hat{f}_\x^{k_1\hdots k_Nl}\right)\right], \label{Fij}
\end{eqnarray}
for all $i$, $j\in \{1,N\}$. Here, the rows of $\mathbf{F}_0$ and each $\mathbf{F}_{ij}$ have the multi-index $k_1\hdots k_Nl$ and the columns $ q_1\hdots q_Nr$. We take all numerical derivatives in \eqref{F0} and \eqref{Fij} via central finite differencing.

Since the diffusion tensor $\mathbf{D}$ is symmetric, for a fixed index $q_1\hdots q_Nr$, $\alpha_{ij}^{q_1\hdots q_Nr} = \alpha_{ji}^{q_1\hdots q_Nr}$ for all $i$, $j\in \{1,N\}$. Hence, the residual in~\eqref{eq:res2} can be written as
\begin{eqnarray}\label{eq:res3}
\boldsymbol{\mathcal{R}}= \mathbf{V} + \mathbf{F}_0\boldsymbol{\alpha}_0 + \sum_i \mathbf{F}_{ii}\boldsymbol{\alpha}_{ii} + \sum_{i<j}(\mathbf{F}_{ij}+\mathbf{F}_{ji})\boldsymbol{\alpha}_{ij}.
\end{eqnarray}
Then the optimization problem over all polynomial coefficients can be written as 
\begin{subequations}\label{opt_fin}
\begin{align}
 \hat{\mathbf{A}} \triangleq
[\hat{\a}_0, \hat{\a}_{11},\hdots, \hat{\a}_{NN}, \hat{\a}_{12}, \hdots, \hat{\a}_{1N}, \hat{\a}_{23}, \hdots, \hat{\a}_{(N-1)N}]^\top
= \argmin_{\mathbf{A}} ||\boldsymbol{\mathcal{R}}(\mathbf{A})||_2^2 + \lam||\mathbf{A}||_1^2,
\end{align}
where
 \begin{align}
 \boldsymbol{\mathcal{R}}(\mathbf{A}) \triangleq \mathbf{V} + [\mathbf{F}_0, \mathbf{F}_{11}, \hdots, \mathbf{F}_{NN}, \mathbf{F}_{12}+ \mathbf{F}_{21}, \hdots, \mathbf{F}_{1N}+\mathbf{F}_{N1}, \mathbf{F}_{23}+\mathbf{F}_{32}, \hdots, \mathbf{F}_{(N-1)N}+\mathbf{F}_{N(N-1)}]
\mathbf{A}.
\end{align}
\end{subequations}

We solve~\eqref{opt_fin} by utilizing the \texttt{lasso} function of MATLAB with a 10-fold CV set on each iteration. We use a mean squared error loss function and a relative tolerance of $10^{-4}$, i.e., the algorithm terminates when the $L^2$ norm of successive estimates of the coefficient vector differ by less than $10^{-4}$. The extrapolation power of our model is tested by fitting the data for the first half of the time horizon, $\left[0,T_f/2\right]$, and analyzing the model performance on the remaining half, $\left(T_f/2,T_f\right]$. 

\begin{remark}
It may be of interest to consider different variations of CV algorithms that are not considered in this study, including least-angle regression and Lasso implemented with Bayes information criterion. Such algorithms may give different results depending on the choice of the hypothesis class. 
\end{remark}


\section{Numerical Experiments} 
\label{sec:applications} 

We consider two applications of the method of distributions and its associated closures to neuron models, namely, the NS-LIF and FHN models, both driven by SW channel noise. The first model has one state variable ($N=1$), and the second has two ($N=2$). SW noise is commonly used in neuroscience, e.g., to model signal transmission~\cite{Kang13}, resonance~\cite{Liang10}, and detection~\cite{Liang11} in nervous systems;  to induce spike death in excitable systems~\cite{guo}; and for weak periodic signal detection in the FHN neuron \cite{yao18}. 

SW noise $\xi(t)$ is a bounded and correlated zero-mean non-Gaussian process that is defined by %
\begin{equation} \label{sw}
\xi(t) \triangleq  \s\sin\left(\sqrt{\frac{2}{\tau}}W_t\right),
\end{equation}
where $\s>0$ is the finite noise amplitude/strength, $\tau>0$ is the correlation length, and  $W_t$ is a standard Wiener process. It has mean $\langle \xi(t)\rangle = 0$ for any fixed time $t$, and covariance function \cite{ning16, yang17, kampen, risken1996}
\begin{equation}
C_\xi(t,s) \triangleq \langle \xi(t) \xi(s) \rangle = \frac{\s^2}{2} \exp\left(- \frac{t-s}{\tau} \right)\left[1 - \exp\left( - \frac{4s}{\tau} \right)\right] \label{cov}, \qquad s\le t.
\end{equation}
For the experiments in Section \ref{LIF_sec}, the RODE~\eqref{ivp2} and all of its corresponding PDF equations are solved up to time $T_f = 50$, while in Section \ref{sec:FHN}, they are solved up to $T_f = 100$.

\subsection{Stochastic NS-LIF model}
\label{LIF_sec}

 LIF neuron models are often used in place of their higher-dimensional counterparts, i.e., models with large~$N$, especially when considering large coupled networks. They describe the basic neuron behavior of slowly charging in the presence of current, spiking after reaching a voltage threshold, and then slowly discharging~\cite{LIF}. In order to present an example where all of the proposed closures are in agreement, the LIF model under consideration is not a spiking model, i.e., it lacks a voltage threshold and resetting mechanism for the membrane potential. We do study spiking dynamics in Section~\ref{sec:FHN} for which there is a discrepancy between the different closures. 
 
In particular, we consider a (one-dimensional) LIF model that is embedded in the (two-dimensional) stochastic FHN model in the excitable regime.\footnote{A FHN neuron is considered to be in the excitable regime when starting in the basin of attraction of a unique stable fixed point, a large excursion, or spike, occurs in phase space and then permanently returns to the fixed point \cite{izhikevich}.} The derivation of the embedding is presented in Section 4 of \cite{yamakou19}. We use their Eq.~4.3, albeit driven by SW noise rather than Gaussian white noise and without a threshold and resetting mechanism, to describe the membrane voltage $x(t) > 0$,
\begin{equation} \label{lif}
\frac{\text dx}{\text dt} = \frac{\s^2}{2x} - \mu x + \xi(t),
\end{equation}
with a given initial condition $x(0) = x_0 \sim |\mathscr{N}(0.1, 0.0125^2)|$ (i.e., a folded Gaussian random variable). In the simulations below, we set $\mu = 0.0312496$ and consider noise amplitudes $\s = 0.0444$, $0.0666$, $0.0888$, and $0.1333$, as in \cite{yamakou19}.  These amplitudes have been appropriately scaled from  $\s = 0.005$, $0.0075$, $0.01$, and $0.015$ in the FHN model. The correlation lengths vary between $\tau = 0.05$, $0.10$, $0.15$, and $0.20$. 

 The RODE~\eqref{lif} is a one-dimensional version of~\eqref{ivp2} with $v(x(t), t; \xi(t)) \equiv \s^2/(2x) - \mu x + \xi$. In phase space, $v(X, t; \xi(t)) \equiv \s^2/(2X) - \mu X + \xi$ is linear in the additive SW noise $\xi(t)$. Hence, $\langle v \rangle \equiv  \s^2/(2X) - \mu X$ and $v' \equiv \xi$. The  terminal problem~\eqref{term1} takes the form
\begin{eqnarray}
 \frac{\text d \chi}{\text ds} = \frac{\s^2}{2\chi} - \mu \chi,  \qquad
 \chi(t) = X, 
 \end{eqnarray}
and admits an analytical solution, yielding a closed-form expression for the associated flow  $\chi(s) = \Phi(s; X,t)$.


\subsubsection{Local LED closure} \label{loc_con}

For the problem under consideration, equations~\eqref{leddrift} and~\eqref{leddiffusion}, which define the drift velocity $\mathscr V$ and the diffusion coefficient $\mathscr D$ in the local LED closure~\eqref{eq:LEDclass}, reduce to
\begin{align}\label{eq:Dprob1}
\mathscr{V}(X,t) \equiv 0 \quad\text{and}\quad
\mathscr{D}(X,t) =  \int_{0}^{t} \J(s; X,t) C_\xi(t,s) \, \text ds,
\end{align}
where 
\begin{equation} \label{eq:Jprob1}
 \J(s; X,t) =  \exp \left(- \int_s^t \frac{\d}{\d\chi} \langle v(\chi(r),r) \rangle \text dr \right) = \exp \left(\int_s^t \left[\frac{\s^2}{2\Phi^2(r; X,t)}+\mu\,\right] \text dr \right).
  \end{equation}
The integral in $\J$ can be evaluated analytically, but this is generally not the case. To be consistent with the approximations that are usually needed, we approximate $\Phi(s; X,t)$ in $\J$ by its terminal condition $X$, resulting in the approximate Jacobian
\begin{equation} \label{jac-lif}
 \tilde{\J}(s; X,t)  = \exp \left(\int_s^t \left[\frac{\s^2}{2X^2}+\mu\,\right] \text dr \right) = \exp \left[ (t-s)\left( \frac{\s^2}{2X^2} + \mu \right) \right].
\end{equation}
With this local LED closure, the single-point PDF $f_x(X;t)$ of the random membrane voltage $x(t)$ satisfies the PDF equation
 \begin{equation}\label{eq:PDFeqProb1}
\frac{ \partial f_x}{\partial t} + \frac{\partial}{\partial X} \left[\left(\frac{\s^2}{2X} - \mu X\right) f_x \right] =  \frac{\partial}{\partial X}  \left(\tilde{\mathscr D}(X,t) \frac{\partial f_x}{\partial X}  \right),
 \end{equation}
 where the diffusion coefficient $\tilde{\mathscr D}$ is computed upon replacing $\J$ with $ \tilde{\J}$ in~\eqref{eq:Dprob1} and analytically evaluating the resulting integral.

\begin{remark}
As mentioned in Remark~\ref{remark:Taylor}, a higher-order Taylor expansion around the terminal condition $(X,t)$ may be used as an alternative approximation for $\Phi(s; X, t)$ in~\eqref{eq:Jprob1}. Our numerical experiments with this approximation showed no improvement in accuracy, but simultaneously induced additional numerical challenges: the resulting linear approximation of the Jacobian determinant $\mathscr J$ behaves like $\exp[(t-s)\Oo(X^{-2})]$, as does $\tilde{\J}$ in~\eqref{jac-lif}, but has singularities at points other than $X=0$.
\end{remark}


\subsubsection{Semi-local LED closure} 
\label{semiloc_con}

\begin{figure}[htbp] 
\centering  
  \includegraphics[width=.75\linewidth]{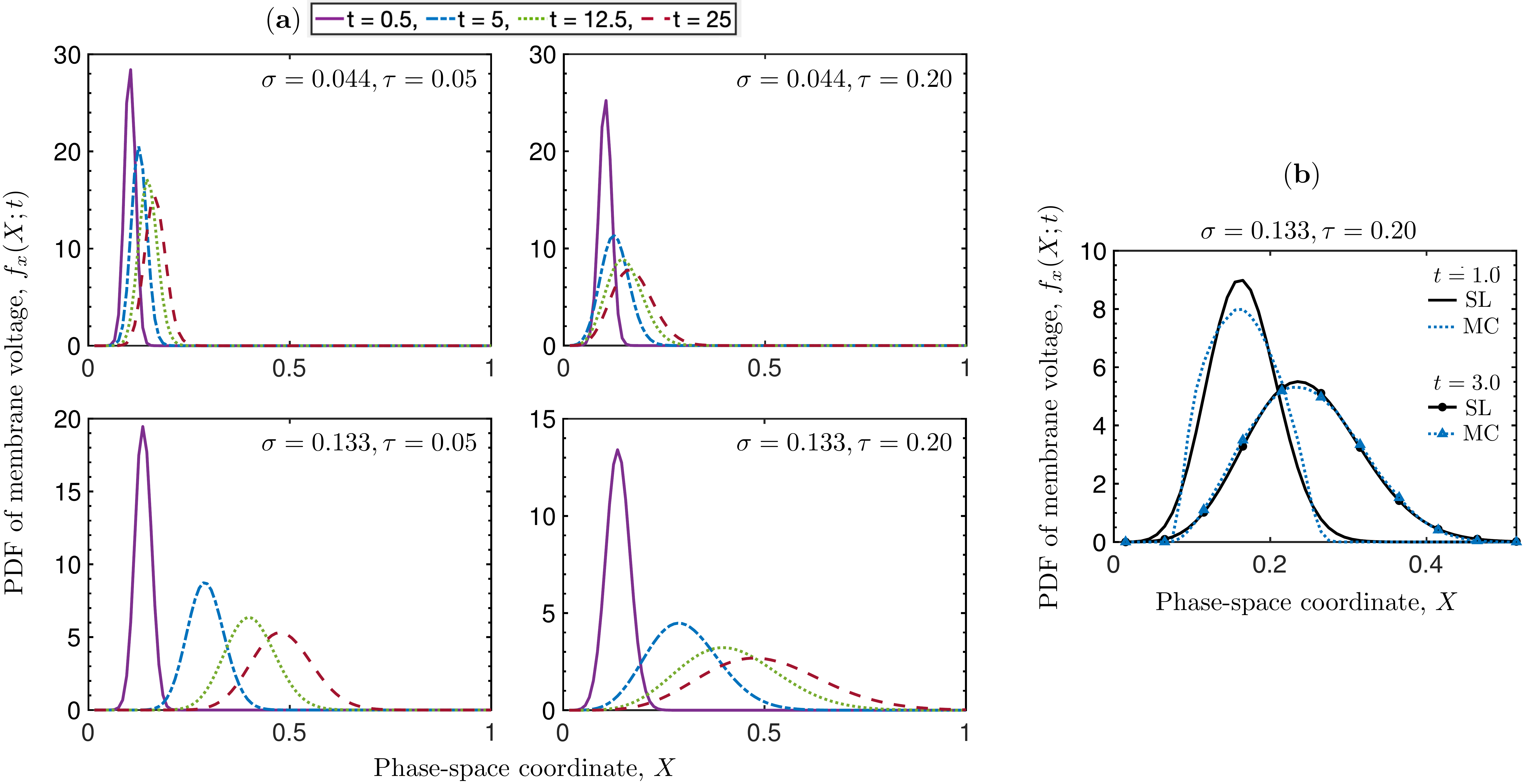}
\caption{ (\textbf{a}) PDF $f_x(X;t)$ of the membrane voltage $x(t)$ in~\eqref{lif}, computed via the PDF equation with the semi-local closure, for several combinations of the strength ($\sigma$) and correlation length ($\tau$) of colored noise $\xi(t)$.
(\textbf{b}) Direct comparison of the PDF $f_x(X;t)$ computed with the semi-local closure and the yardstick PDF $f_\text{MC}(X;t)$ for $\s=0.133$ and $\tau=0.20$ at times $t = 1$ and $3$.} 
\label{fig:sloc1}
\end{figure}

\begin{figure}[htbp] 
\centering  
\includegraphics[width=.7\linewidth]{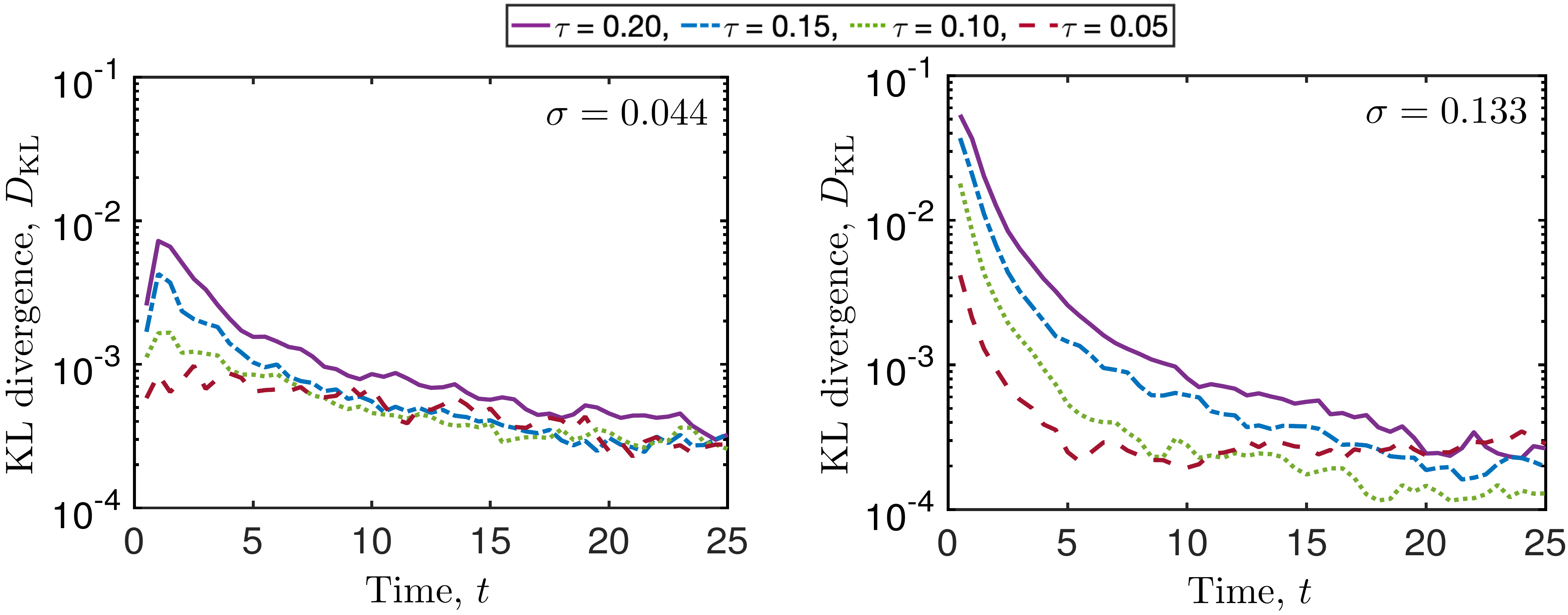}
\caption{Temporal evolution of the KL divergence, $D_\text{KL}(f_\text{MC}||f_x)$, between the membrane voltage PDF computed with the semi-local closure, $f_x(X;t)$, and its Monte Carlo counterpart, $f_\text{MC}(X;t)$, for several combinations of the strength ($\sigma$) and correlation length ($\tau$) of colored noise $\xi(t)$. }
\label{fig:sloc1_err}
\end{figure}

For the problem under consideration, the semi-local closure~\eqref{ledmod2} reduces to 
\begin{align} \label{sloc1}
\langle v'\Pi'\rangle(X,t) \approx - D_2(t) \frac{\partial f_x}{\partial X},
\end{align}
where the diffusion coefficient $D_2(t)$ is computed analytically,
\begin{align}\label{eq:D2}
D_2(t) = \int_{0}^{t} \tilde{\J}(s ; X,t) \exp\left((t-s)\frac{\text d \langle v(X,t)\rangle }{\text dX} \right) C_\xi(t,s) \text ds= \int_{0}^{t} C_\xi(t,s) \text ds.
\end{align}
The resulting PDF equation for $f_x(X;t)$ is identical to~\eqref{eq:PDFeqProb1}, except that now $D_2(t)$ is used in place of $\tilde{\mathscr D}(X,t)$.\footnote{We found the semi-local closure~\eqref{ledmod1} to provide only minor improvements in accuracy relative to the closure~\eqref{ledmod2}, while being significantly more involved computationally. Consequently, only the numerical results for the semi-local closure~\eqref{ledmod2} are presented therein.} A solution to this equation, $f_x(X;t)$, is shown in Fig.~\ref{fig:sloc1}a for several combinations of the meta-parameters $\sigma$ and $\tau$ characterizing the strength and correlation length of the colored noise $\xi(t)$, respectively. For the stochastic NS-LIF model~\eqref{lif}, the PDFs of membrane voltage, $f_x(X;t)$, computed via the PDF equation with the nonlocal and local closures are visually indistinguishable from the PDF with the semi-local closure shown in Fig.~\ref{fig:sloc1}a. Fig.~\ref{fig:sloc1}b gives a direct comparison of the PDF computed with the semi-local closure and the yardstick PDF, $f_\text{MC}(X;t)$, estimated via high-resolution Monte Carlo simulations.  The latter required $N_\text{MC} \approx 1.5 \times 10^5$ realizations to converge. The results in Fig.~\ref{fig:sloc1}b are displayed for the worst case scenario, which occurs for the largest choices of $\s$ and $\tau$ and at early times, as seen by the error plot in Fig.~\ref{fig:sloc1_err}.

Fig.~\ref{fig:sloc1_err} provides a quantitative assessment of the accuracy of the semi-local closure. It displays the temporal evolution of the KL divergence $D_\text{KL}(f_\text{MC}||f_x )$ between $f_x(X;t)$ and $f_\text{MC}(X;t)$ for the noise strength $\sigma = 0.044$ and $0.133$. (The errors associated with the intermediate values of $\sigma = 0.066$ and $0.088$ follows the same pattern and, hence, are omitted from Fig.~\ref{fig:sloc1_err}). The KL divergence increases with $\sigma$ and $\tau$, as to be  expected from the theoretic considerations, since the LED-based closures can be thought of as leading-order terms in perturbation expansions in the powers of $\sigma$ and $\tau$. 
Regardless of the choice of the meta-parameter values, $D_\text{KL}(f_\text{MC}||f_x )$ generally decreases with time $t$. That is because the membrane voltage PDF approaches its nearly Gaussian steady state, wherein the PDF and its derivatives with respect to $X$ are almost constant over the correlation-length intervals $(t-\tau,t)$, an approximation that underpins both the local and semi-local closures. For all choices of $\sigma$ and $\tau$ and for all time $t$, the nonlocal, semi-local, and local closures yield PDFs $f_x(X;t)$ within a KL divergence of $1.6\times 10^{-4}$ of one another, with the largest error occurring for the largest values of $\sigma$ and $\tau$.\footnote{Although the accuracy of the local and semi-local closures is nearly identical for the stochastic NS-LIF model~\eqref{lif}, this is generally not the case, as seen in Section \ref{sec:FHN}. }
The local and semi-local approaches are four orders of magnitude faster than the Monte Carlo approach.

\subsubsection{Data-driven closure} 
\label{sec:dd1}

\begin{figure}[htbp] 
\centering  
\includegraphics[width=.7\linewidth]{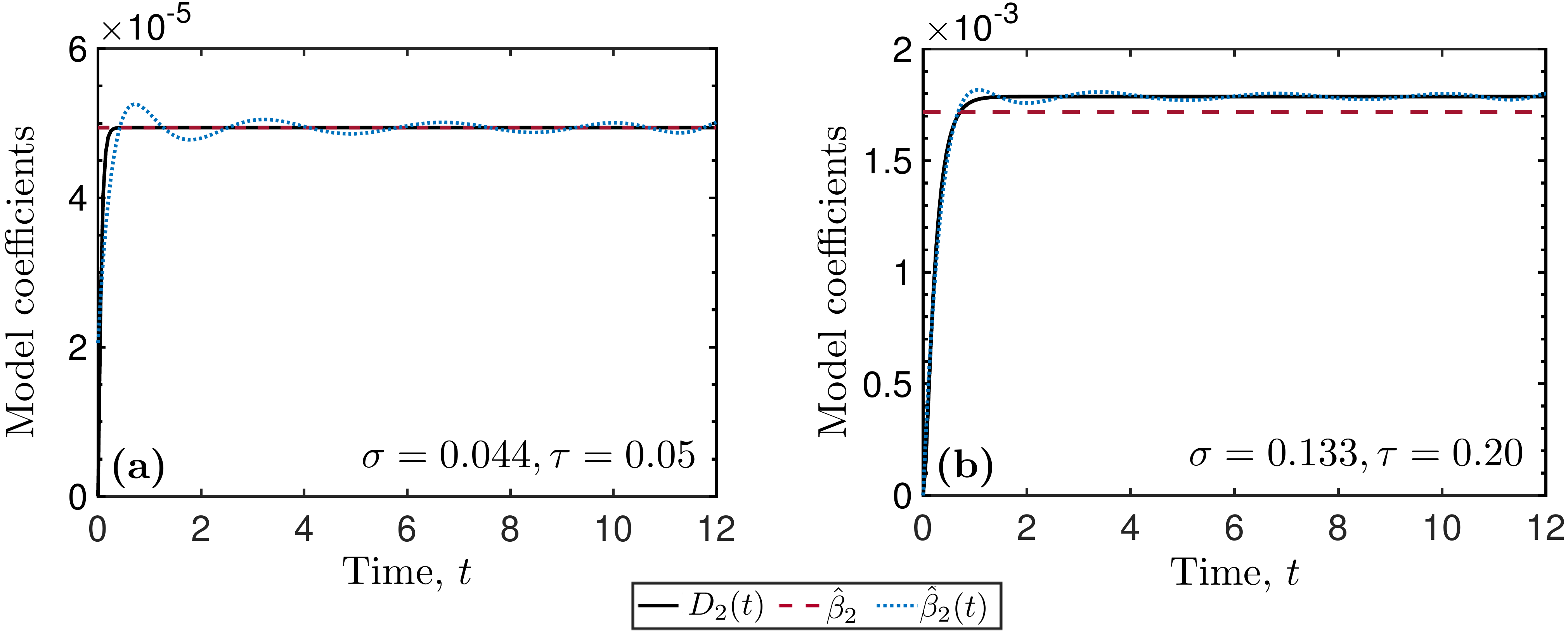}
\caption{Diffusion coefficients associated with the semi-local closure, $D_2(t)$, the data-driven closure with constant coefficients, $\hat \beta_2$, and the data-driven closure with a temporal Legendre basis expansion, $\hat \beta_2(t)$. We present the cases for which the KL divergence $D_\text{KL}(f_\text{MC} \| f_x)$ of the data-driven closure with constant $\hat\b$ is highest. In all cases, the data-driven coefficients approximated with a Legendre basis expansion agree well with the other closures.}
\label{fig:coeff1}
\end{figure}

Since the diffusion coefficient for the semi-local closure, $D_2$, depends only on time, see~\eqref{eq:D2}, we consider spatially independent variable coefficients $\b \equiv \b(t)$ in the data-driven closure~\eqref{cheb}, significantly reducing the dimensionality of the optimization problem~\eqref{min1}\footnote{To be concrete, we represented $\beta_i(t)$ ($i=1,2$) with the first ten Legendre polynomials.}. Furthermore, since $D_2(t)$ is a monotonically increasing function that quickly reaches an asymptote (Fig.~\ref{fig:coeff1}), we also investigate a data-driven closure with constant coefficients $\b$, eliminating the need for a polynomial expansion and further simplifying the optimization problem. In both cases, the one-dimensional version of~\eqref{eq:ML} gives rise to the hypothesis 
\begin{equation} \label{dd_cov1}
 \frac{\partial \langle v'\Pi'\rangle }{\partial X} \approx \beta_1 \frac{\partial f_x}{\partial X} + \beta_2 \frac{\partial^2 f_x}{\partial X^2}, 
 \end{equation}
and either constant or time-dependent coefficients $\b = (\beta_1,\beta_2)^\top$ are found by sparse regression~\eqref{min1} in which the hyper-parameter $\lambda$ is chosen to minimize the mean squared error. 

Regardless of whether the coefficients $\b$ are treated as constant or time-dependent, the sparse regression yields $\beta_1 \equiv0$, so that $\beta_2$ becomes an approximation of the diffusion coefficient $D_2(t)$; the optimal estimates of both constant and time-varying $\beta_2$, denoted by $\hat{\cdot}$, are compared with $D_2(t)$ in Fig.~\ref{fig:coeff1}. For small values of $\s$ and $\tau$ (Fig.~\ref{fig:coeff1}a), $D_2(t)$ quickly approaches its asymptote, rendering its approximation with the constant $\hat\b_2$ accurate. Meanwhile, the time-dependent approximation $\hat\b_2(t)$ suffers from oscillations around the asymptote, with a noticeable overshoot near $t=1$. This discrepancy becomes less pronounced as the number of polynomial bases is increased. As $\s$ and $\tau$ increase (Fig.~\ref{fig:coeff1}b), the rate at which $D_2(t)$ reaches its asymptote decreases.  This causes the constant approximation $\hat\beta_2$ to slightly underestimate $D_2(t)$ because it has to ``average out" low levels of diffusion at early times. However, this slower transition benefits the time-dependent approximation $\hat\b_2(t)$, allowing for fewer polynomial bases and less pronounced oscillations and overshoot. 

\begin{figure}[htbp] 
\centering  
\includegraphics[width=.7\linewidth]{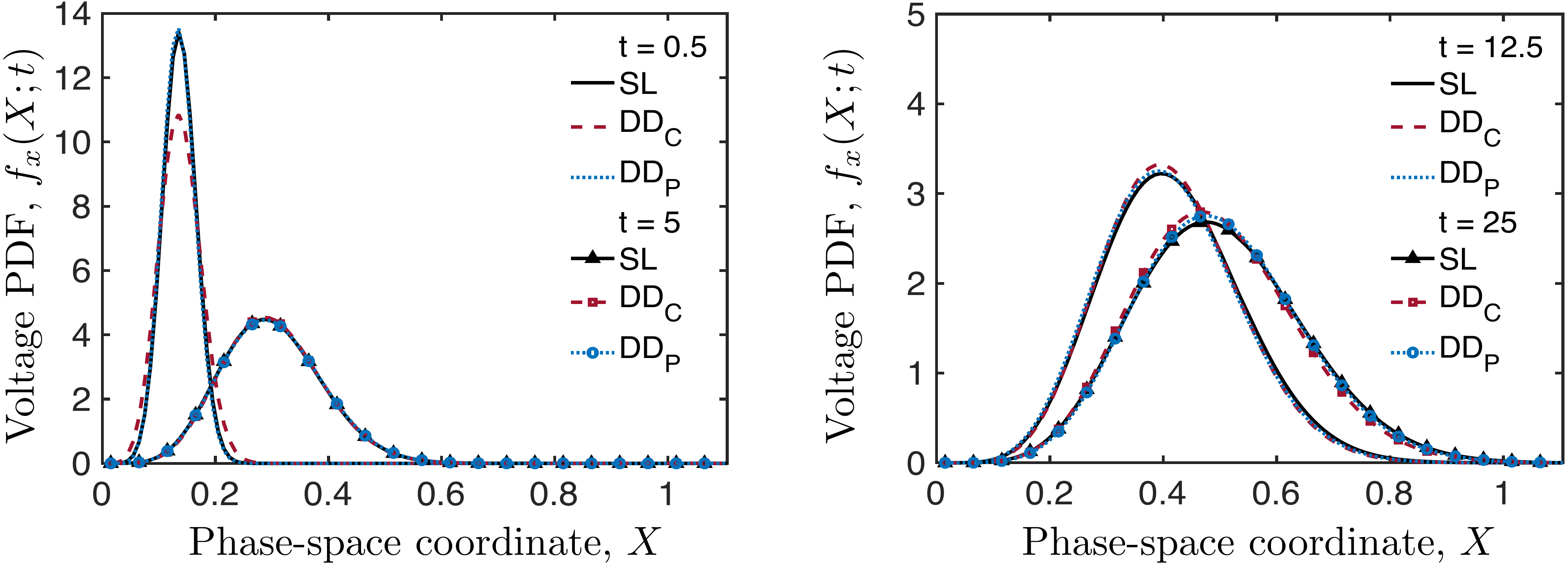}
\caption{ Temporal snapshots (at times $t = 0.5$, $5$, $12.5$, and $25$; for $\s = 0.133$ and $\tau = 0.20$) of the membrane voltage PDFs $f_x(X;t)$ alternatively computed with the semi-local closure (SL), the data-driven closure with constant coefficients ($\text{DD}_{\text{C}}$), and the data-driven closure with a temporal Legendre basis expansion ($\text{DD}_{\text{P}}$).}
\label{fig:mc_sloc_lasso1}
\end{figure}

Fig.~\ref{fig:mc_sloc_lasso1} reveals that, while the semi-local closure outperforms the data-driven closure (it is visually indistinguishable from $f_\text{MC}$), the data-driven closures perform remarkably well. For large values of $\s$ and $\tau$, allowing $\hat \beta_2$ to vary in time improves the accuracy of the data-driven closure at early times, but gives only minor improvements at later times. To a certain extent, the under-approximation of $D_2(t)$ by the constant  $\hat \beta_2$ at later times is balanced by its over-approximation of $D_2(t)$ at early times. However, the discrepancy does increase at later extrapolated times, as the constant $\hat \beta_2$ leads to the data-driven PDF $f_x(X;t)$ being under-diffused. This comparison is provided for the largest values of $\sigma$ and $\tau$ considered in our study ($\sigma = 0.133$ and $\tau = 0.20$). 

\begin{figure}[htbp] 
\centering  
   \includegraphics[width=.8\linewidth]{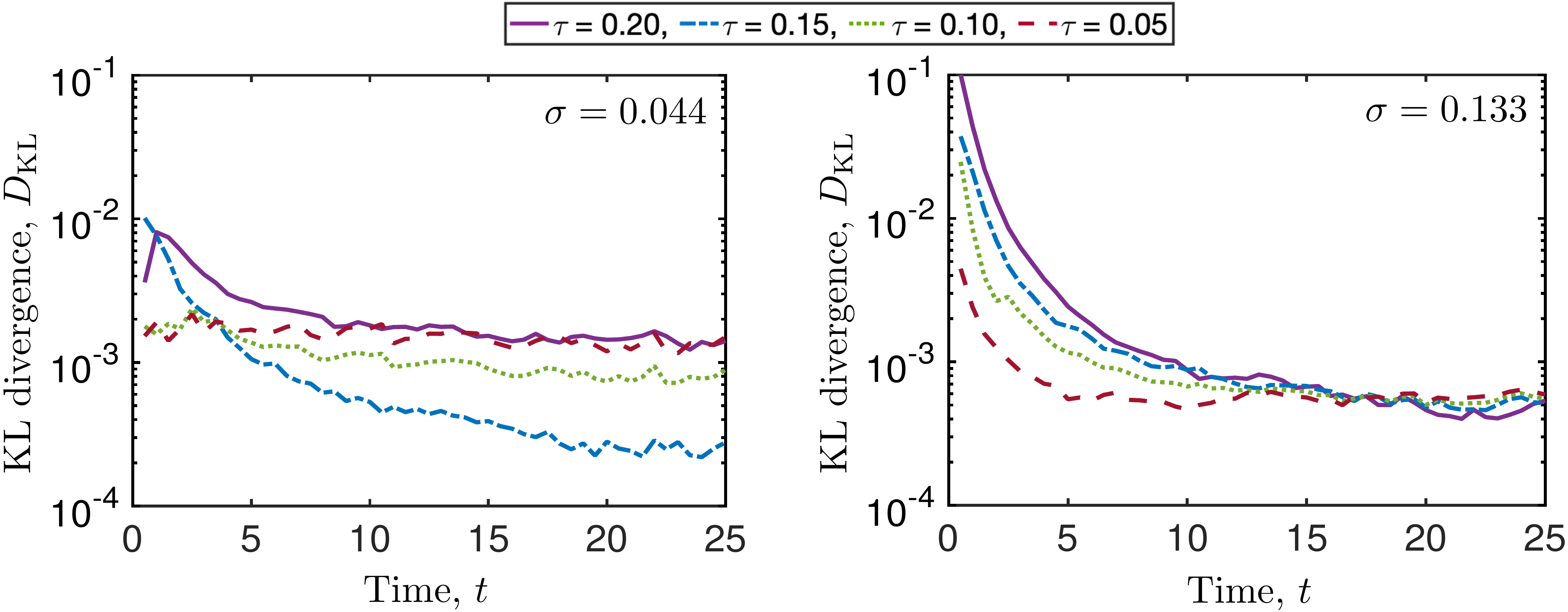}
\caption{Temporal evolution of the KL divergence, $D_\text{KL}(f_\text{MC}||f_x)$, between the membrane voltage PDF computed with the data-driven closure, $f_x(X;t)$, and its Monte Carlo counterpart, $f_\text{MC}(X;t)$, for selected combinations of the strength ($\sigma$) and correlation length ($\tau$) of colored noise $\xi(t)$. The data array $\hat{f}_x$ used in the data-driven closure is computed with $N_\text{MC}^\text{tr} = 3\times 10^4$ Monte Carlo realizations.}
\label{fig:las_sloc1_err}
\end{figure}

Fig.~\ref{fig:las_sloc1_err} elaborates upon this visual observations by presenting the temporal evolution of the KL divergence $D_\text{KL}(f_\text{MC}||f_x )$ between the ``exact'' solution $f_\text{MC}(X;t)$ and its approximations $f_x(X;t)$ obtained with the data-driven closure. We illustrate the performance of the least computationally expensive data-driven closure, i.e., the one that relies on the constant coefficients $\hat\b\in\mathbb{R}^2$.  As $\s$ and $\tau$ decrease, the discrepancy between the the semi-local closure and the data-driven closure increases, with the largest difference in their KL divergences occurring for $\s=0.044$ and $\tau = 0.05$ at later times. At first glance, this may seem contradictory to Fig.~\ref{fig:coeff1}, but one must consider that the diffusion coefficients decrease with $\s$ and $\tau$, and that $\s$ also influences advection as it is included in the drift term in~\eqref{lif}. When $\s$ and $\tau$ are small, the resulting PDF $f_x(X;t)$ advects and diffuses at a slow rate (see Fig.~\ref{fig:sloc1}), resulting in a spikier PDF.  Hence, in terms of relative entropy, i.e., the KL divergence, $f_x(X;t)$ is more sensitive to small changes in the diffusion coefficient. Unlike in the LED theory, the data-driven closure does not guarantee that $f_x(X;t)$ should improve as $\s$ and $\tau$ decrease. Regardless of these observations, the data-driven closure remains accurate when compared to the semi-local closure.

\begin{figure}[htbp] 
\centering  
    \includegraphics[width=.7\linewidth]{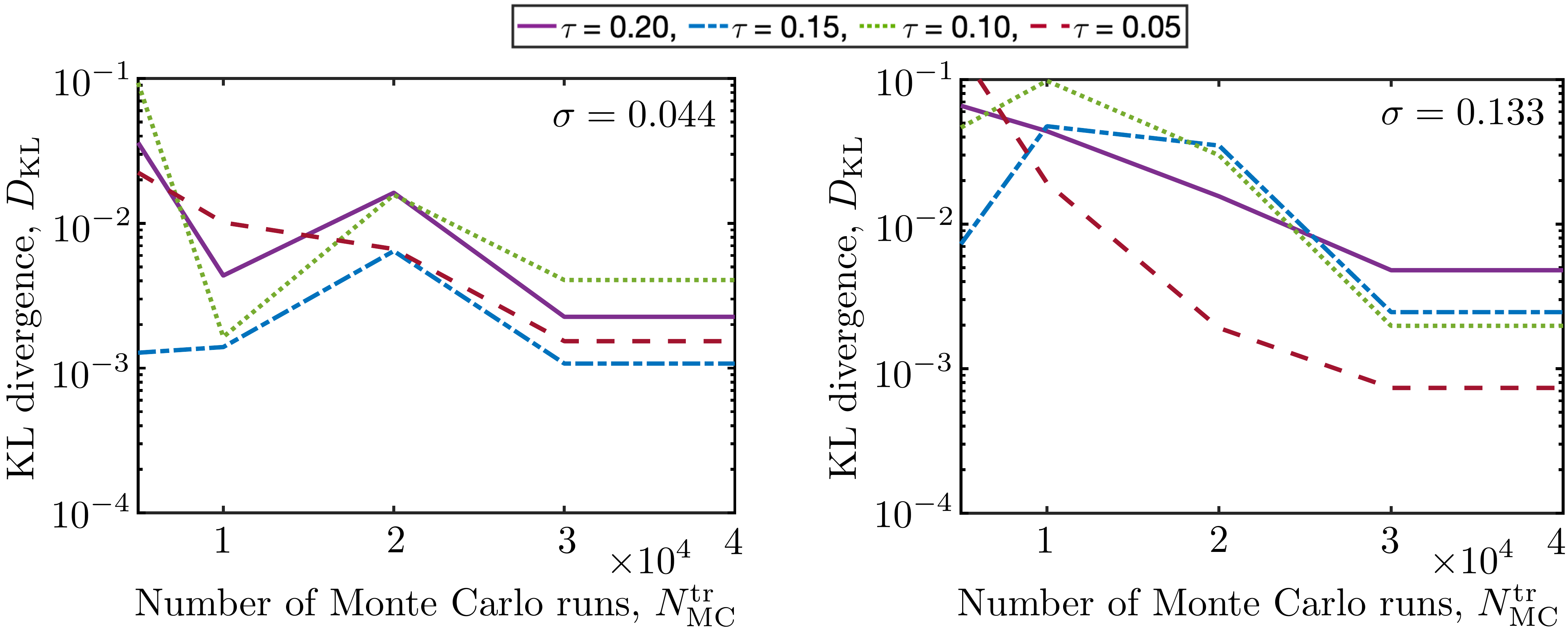}
\caption{Dependence of the KL divergence $D_\text{KL}(f_\text{MC} \| f_x)$, averaged over all temporal grid nodes, on the number of Monte Carlo runs used in the sparse regression, $N_\text{MC}^\text{tr}$. }
\label{fig:mc1_err}
\end{figure}

The residual $\mathcal R$ in the optimization problem~\eqref{min1} uses the data array $\hat{f_x}$ computed from $N_\text{MC}^\text{tr}$ Monte Carlo realizations. Hence, the optimal values of $\b$ and, hence, the PDF $f_x(X;t)$ computed with~\eqref{split1},~\eqref{split2} and~\eqref{dd_cov1} are affected by the choice of $N_\text{MC}^\text{tr}$. Fig.~\ref{fig:mc1_err} sheds light on this dependence by exhibiting the KL divergence $D_\text{KL}(f_\text{MC}||f_x )$, averaged over all temporal grid nodes, as function of $N_\text{MC}^\text{tr}$. For all choices of $\sigma$ and $\tau$ considered, the optimal solution $\hat\b$ to~\eqref{min1} needs $N_\text{MC}^\text{tr} \approx 3\times 10^4$ Monte Carlo trials to stabilize, and the accuracy of the data-driven closure does not improve for $N_\text{MC}^\text{tr} >  3\times 10^4$. This is less than $20\%$ of the number of realizations, $N_\text{MC}$, needed to compute  $f_\text{MC}$.  The data-driven results presented in Figs.~\ref{fig:coeff1}--\ref{fig:las_sloc1_err} are obtained with $N_\text{MC}^\text{tr} =  3\times 10^4$ Monte Carlo trials.

\subsection{Stochastic FHN model} \label{sec:FHN}

The FHN neuron is controlled by two state variables, the membrane potential $x_1(t)$ and the recovery variable $x_2(t)$ that restores a neuron's resting state. In the excitable regime, its dynamics is described by
   \begin{eqnarray}
     \begin{bmatrix}
         \dot x_1 \\
         \dot x_2 \\
        \end{bmatrix}
        = \begin{bmatrix}
        x_1 - x_1^3 / 3 - x_2 + I \\
         \epsilon(x_1+\alpha -\beta x_2) +\xi\\
        \end{bmatrix}\label{fhn1}
 \end{eqnarray}
where SW channel noise  $\xi(t)$ is defined by~\eqref{sw}; $I$ is the external input current; $\epsilon$ is a small timescale parameter that ensures that the dynamics of $x_1$ are much faster than $x_2$; and $\alpha$ and $\beta$ are model parameters. In the simulations reported below, we set $I = 0.265$, $\epsilon=0.08$, $\alpha = 0.7$ and $\beta = 0.75$, in accordance with~\cite{yamakou19}. With these parameter values, the deterministic version of~\ref{fhn1}, i.e., in the absence of the noise $\xi$, is in the excitable regime\footnote{A neuron is considered to be in the excitable regime when starting in the basin of attraction of a unique stable fixed point, a large excursion, or spike, occurs in the phase space and then permanently returns to the fixed point \cite{izhikevich}.} and has a unique stable fixed point at $(x_1,x_2) = (-1.00125, -0.401665)$. In the stochastic setting, the channel noise $\xi(t)$ may perturb the trajectory away from the fixed point, causing repeated firing.  The frequency of this repeated firing increases with the noise amplitude $\s$~\cite{yamakou19}. In our simulations we consider noise amplitudes of $\s = 0.01$, $0.05$, and $0.2$, and correlation lengths $\tau = 0.01$, $0.1$, and $5.0$. The initial state $\x(0) = \x_0$ is an uncorrelated joint Gaussian whose mean is perturbed from the stable fixed point:
   \begin{eqnarray*}
\x_0&\sim \mathscr{N} \left(  \begin{bmatrix}
         -1.00125  \\
         -0.411665 \\
         \end{bmatrix}
         ,\begin{bmatrix}
	0.01 & 0 \\
	 0 & 0.01 \\
        \end{bmatrix}\right).
\end{eqnarray*}
  
The approximate Jacobian determinant $\tilde{\J}$ associated with the FHN model (\ref{fhn1}) needed for both the local and semi-local LED closures is given by 
 \begin{equation*}
   \tilde{\J}(s; \X,t) = \exp(-(t-s)(1-\epsilon\beta-X_1^2)).
  \end{equation*}
  
  
  \subsubsection{Local LED closure}
  
For the problem under consideration, equations~\eqref{leddrift} and~\eqref{leddiffusion} with $\J \approx \tilde{\J}$ yield analytical expressions for the drift velocity $\mathscr V$ and diffusion coefficient $\mathscr D$, 
\begin{subequations}
\begin{align}
\mathscr{V}(\X,t) \equiv \;&
     \begin{bmatrix}
         0 \\
         0 \\
       \end{bmatrix}, \\
\mathscr{D}(\X,t) \triangleq \;&  \int_{0}^{t} \tilde{\J}(s; \X,t) \langle\v'(\X,t)\v'^\top(\boldsymbol\Phi(s; \X,t),s)\rangle \text ds = D_3(X_1,t) 
\begin{bmatrix}
	0 & 0 \\
	 0 & 1 \\
        \end{bmatrix},
\end{align}
where
 \begin{align}
D_3(X_1,t) &=   \int_{0}^{t} \tilde{\J}(s; \X,t)C_\xi(t,s) \text ds.
\label{loc2}
\end{align}
\end{subequations}
The integral in~\eqref{loc2} is computed analytically.  The PDF equation~\eqref{ledclassic} for the joint density $f_\x(\X;t)$ simplifies to
 \begin{align}
 \frac{\partial f_\x}{\partial t} + \nabla_\X \cdot   \left(
 \begin{bmatrix}  X_1 - X_1^3/3 - X_2 + I \\ 
  \epsilon(X_1+\alpha -\beta X_2) \\
  \end{bmatrix} 
  f_\x \right)= D_3\frac{\partial^2 f_\x}{\partial X_2^2}.
 \end{align}

\begin{figure}[htbp] 
\centering  
\includegraphics[width=.7\linewidth]{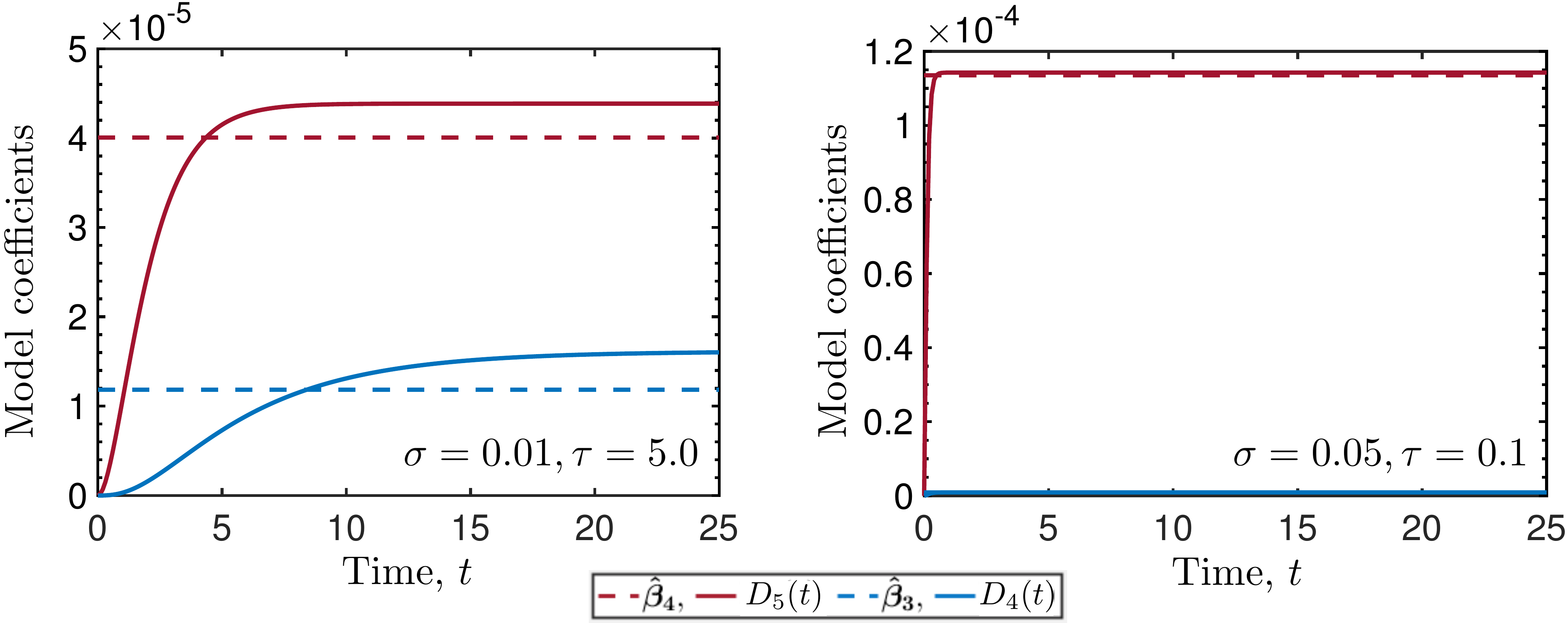}
\caption{The coefficients of the semi-local closure, $D_4$ and $D_5$, and the data-driven closure, $\hat\beta_3$ and $\hat\beta_4$. For $\sigma = 0.05$ and $\tau = 0.1$, the optimization algorithm sets $\hat\beta_3=0$ while $D_4 = \mathcal O(10^{-7})$, which is near zero when compared with $\hat\beta_4$ and $D_5$.}
\label{fig:coeff2}
\end{figure}

  
 \subsubsection{Semi-local LED closure} \label{sec:semiloc2}
  
\begin{figure}[htbp]
\centering
  \includegraphics[width=.7\linewidth]{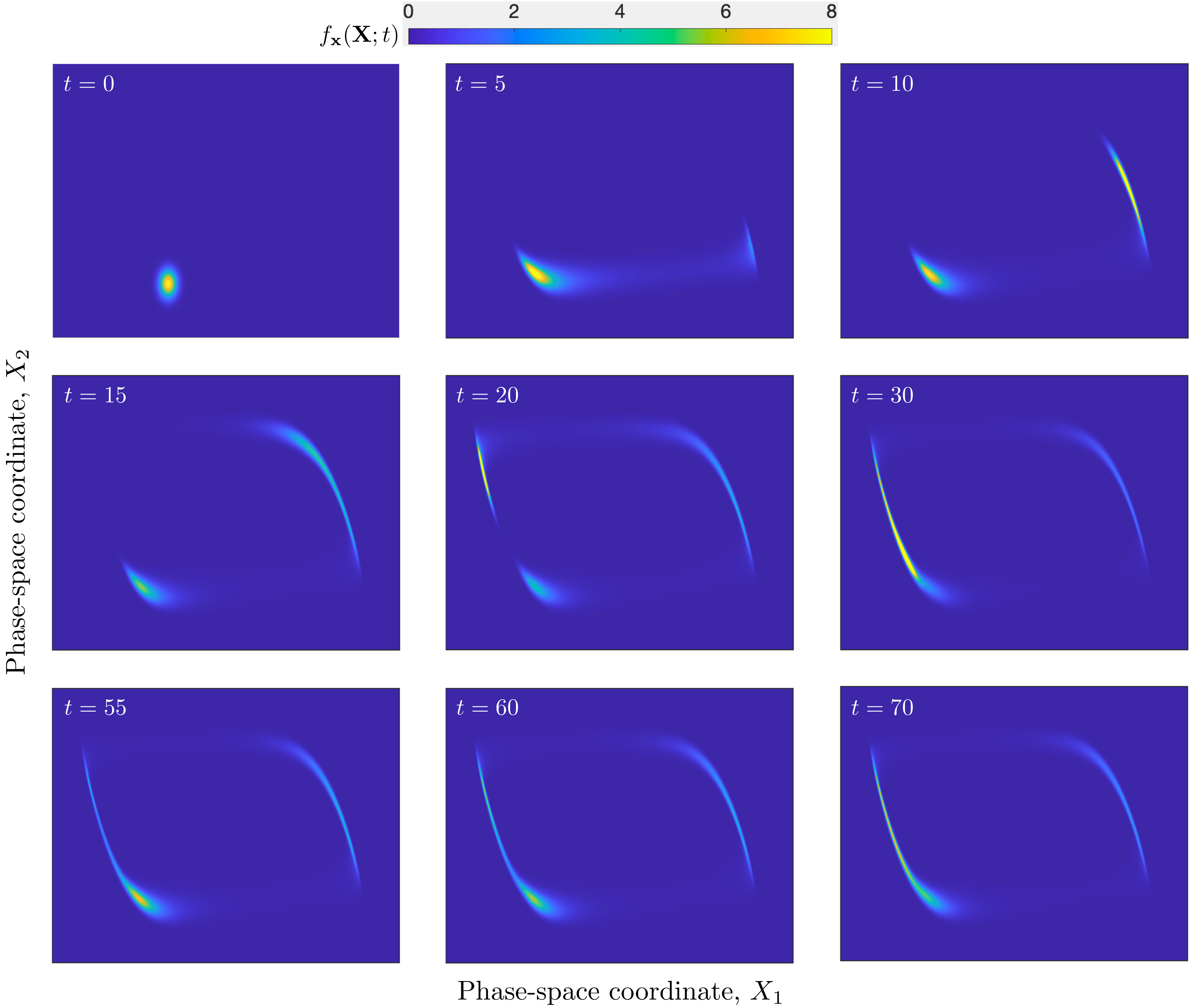}
\caption{Temporal snapshots of the joint  PDF $f_\x(\X;t)$ of the random state variables $x_1(t)$ and $x_2(t)$ in the FHN model~\eqref{fhn1}. The dynamics of $f_\x(\X;t)$ is governed by the PDF equation~\eqref{eq:sl2d}, for $\s = 0.2$ and $\tau = 0.1$. }
\label{fig:fhn_plots}
\end{figure}

 The semi-local closure~\eqref{ledmod2} takes the form
\begin{eqnarray}
\langle\v'\Pi'\rangle(\X,t) = - \int_{0}^{t} \tilde{\J}(s ; \X,t)C_\xi(t,s) \exp((t-s)\j(\X,t))^\top
\begin{bmatrix}
	0 & 0 \\
	 0 & 1 \\
        \end{bmatrix}  \grad_{\X} \,f_\x(\X;t) \text ds.
\label{semlocdiff}
\end{eqnarray}
The integrals in (\ref{semlocdiff}) are computed exactly with a symbolic computation software  \texttt{Mathematica}, giving rise to the PDF equation
 \begin{equation}\label{eq:sl2d}
  \frac{\partial f_\x}{\partial t} + \nabla_\X \cdot   \left(
 \begin{bmatrix}  X_1 - X_1^3/3 - X_2 + I \\ 
  \epsilon(X_1+\alpha -\beta X_2) \\
  \end{bmatrix} 
  f_\x \right) = \frac{\partial }{\partial X_1} \left(D_4 \frac{\partial f_\x}{\partial X_2} \right) + D_5 \frac{\partial^2 f_\x}{\partial X_2^2}.
 \end{equation}
where $D_4(X_1,t)$ and $D_5(X_1,t)$ are known functions whose explicit expressions are omitted for brevity, and whose graphical representation is provided in Fig.~\ref{fig:coeff2}. (The $X_1$-dependence of these coefficients is insignificant, and $D_4(\cdot, t)$ and $D_5(\cdot,t)$ are plotted for $X_1=0$.) Fig.~\ref{fig:fhn_plots} shows temporal snapshots of the joint PDF $f_\x(\X;\cdot)$ computed with~\eqref{eq:sl2d}. The shape of this PDF reveals that the ensemble mean $\langle \x(t) \rangle$ in and of itself does not accurately capture the dynamics of~\eqref{fhn1}.

Fig.~\ref{fig:fhn_err} depicts the temporal evolution of the KL divergence $D_\text{KL}(f_\text{MC} \| f_\x)$. It serves to ascertain the accuracy of the semi-local closure that underpins the PDF equation~\eqref{eq:sl2d}.  As expected, the accuracy increases as the correlation length $\tau$ decreases. As the noise amplitude $\s$ decreases, the different choices for $\tau$ become less impactful as the errors are more tightly clustered. Although not shown here, we found that, for all choices of $\s$, the local closure agrees with the semi-local closure within a KL divergence of $10^{-4}$ for $\tau = 0.01$. For $\tau = 0.1$, the KL divergence of the local closure is between $0.1$ and $0.5$ more than that of the semi-local closure; and for $\tau = 5.0$, it is up to a full order of magnitude more than that of the semi-local closure.


\subsubsection{Data-driven closure} 
\label{sec:dd2}

Guided by the results reported in Section~\ref{sec:dd1}, we focus on constant coefficients $\b$ so that the data-driven closure~\eqref{eq:ML} gives rise to a hypothesis
\begin{equation} \label{dd_fhn_cov1}
 \nabla_\X\cdot \langle \mathbf v'\Pi'\rangle = \beta_1 \left(\frac{\partial f_\x }{ \partial X_1} + \frac{\partial f_\x}{\partial X_2}\right) + \beta_2 \frac{\partial^2 f_\x}{\partial X_1^2} + \beta_3 \frac{\partial^2 f_\x}{\partial X_1 \partial X_2} + \beta_4 \frac{\partial^2 f_\x }{ \partial X_2^2}.
 \end{equation}
The sparse regression optimization~\eqref{min1} identifies $\hat\beta_1=\hat\beta_2=0$, so that the resulting PDF equation~\eqref{pdfeq1} matches the PDF equation~\eqref{eq:sl2d} obtained with the semi-local closure, with the constants $\hat\beta_3$ and $\hat\beta_4$ serving as empirical representations of $D_4(X_1,t)$ and $D_5(X_1,t)$, respectively.  The values of $\hat\beta_3$ and $\hat\beta_4$ are shown in Fig.~\ref{fig:coeff2} for two combinations of $\sigma$ and $\tau$. These values slightly underestimate the asymptotes of the semi-local diffusion coefficients $D_4(\cdot,t)$ and $D_5(\cdot,t)$ due to ``averaging out'' of low levels of diffusion at early times. 

\begin{figure}[htbp]
\centering
 \includegraphics[width=0.9\linewidth]{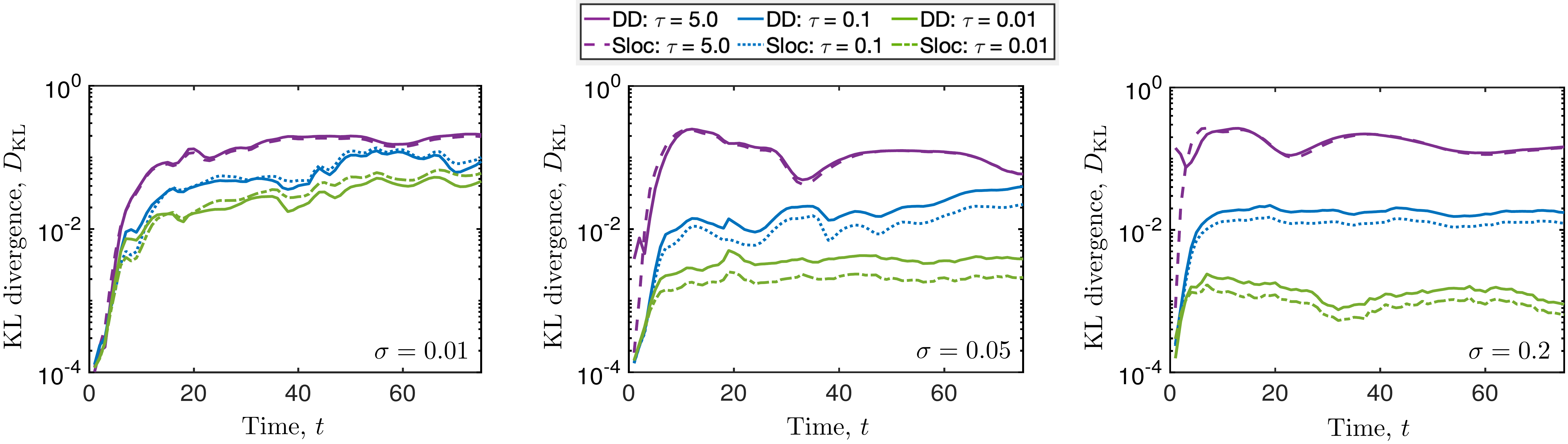}
\caption{Temporal evolution of the KL divergence $D_\text{KL}(f_\text{MC} \| f_\x)$ between the yardstick Monte Carlo solution $f_\text{MC}(\X;t)$ of~\eqref{fhn1} and the PDF $f_\x(\X;t)$ computed, alternatively, via~\eqref{eq:sl2d} and~\eqref{pdfeq1} with the data-driven closure~\eqref{dd_fhn_cov1}. }
\label{fig:fhn_err}
\end{figure}

This discrepancy gives rise to the relatively small difference in predictions of $f_\x(\X;t)$, as quantified by the KL divergence $D_\text{KL}(f_\text{MC} \| f_\x)$ in Fig.~\ref{fig:fhn_err}. The semi-local closure outperforms the data-driven closure, but just barely. The close agreement between the PDF predictions based on these two closures is remarkably robust, holding for all choices of $\sigma$ and $\tau$ considered. Although not shown here, both closures are considerably more accurate than the local closure for larger $\tau$, as discussed in Section~\ref{sec:semiloc2}. 

\begin{figure}[htbp]
\centering
   \includegraphics[width=.8\linewidth]{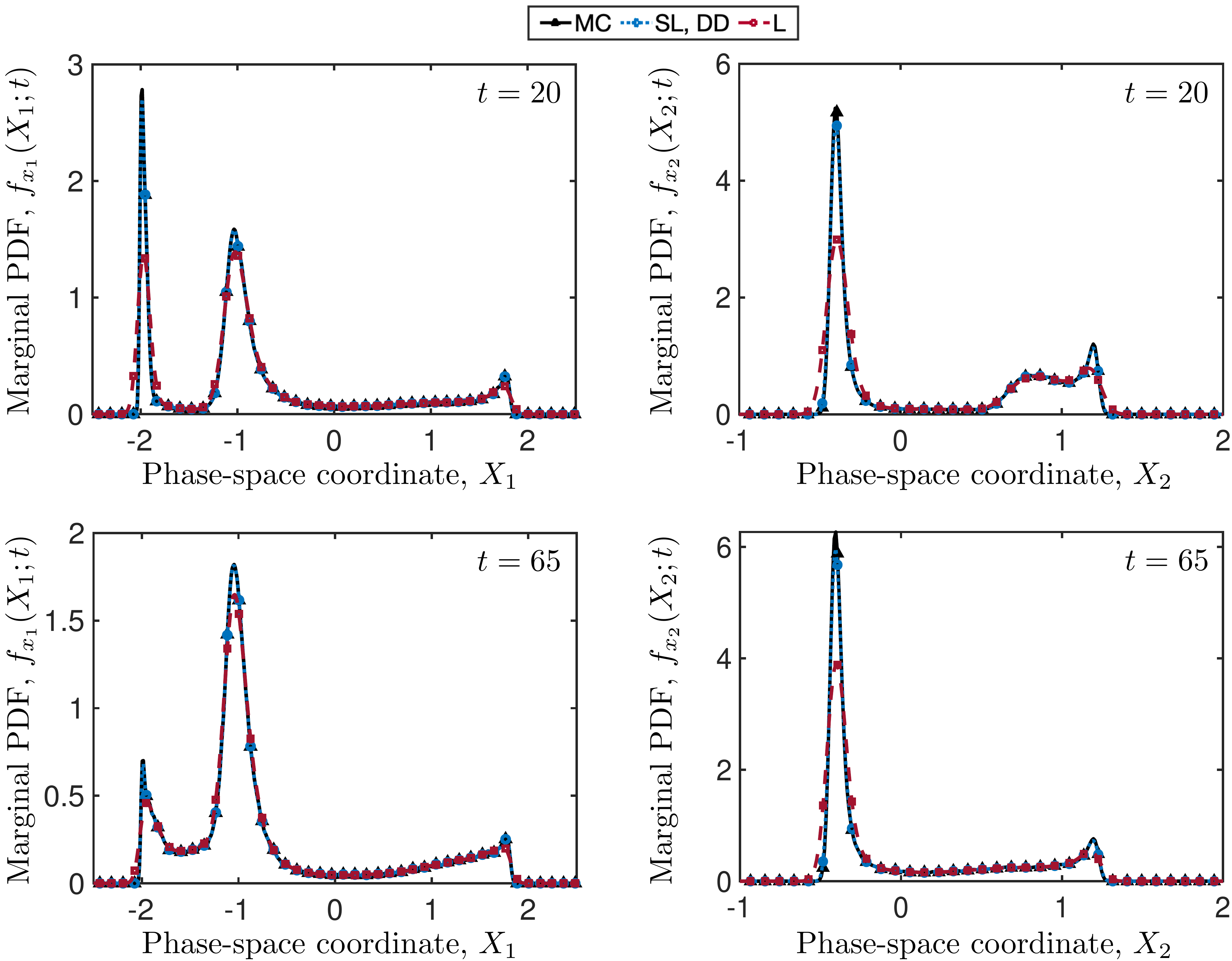}
\caption{Snapshots (at times $t= 20$ and $65$) of the marginal PDFs $f_{x_1}(X_1;t)$ (left column) and $f_{x_2}(X_2;t)$ (right column) alternatively computed with the local (L), semi-local (SL) and data-driven (DD) closures and with Monte Carlo simulations (MC), for the stochastic FHN neuron with  $\s = 0.05$ and $\tau = 0.1$.}
\label{fig:fhn_marg}
\end{figure}

The close agreement between predictions of the joint PDF $f_\x(\X;t)$ based on the semi-local and data-driven closures translates into the close agreement between predictions of the corresponding marginal PDFs $f_{x_1}(X_1;t)$ and $f_{x_2}(X_2;t)$ (Fig.~\ref{fig:fhn_marg}). The semi-local and data-driven closures accurately capture all features of the Monte Carlo solution.  While capturing the coarse dynamics of the Monte Carlo solution, the local closure is too diffusive to accurately capture the dynamics with sharp peaks. Since the PDFs associated with the semi-local and data-driven closures are visually indistinguishable, we include them as the same line-plot.

Finally, we note that an optimal solution $\hat\b(N_\text{MC}^\text{tr})$ to the sparse regression problem~\eqref{min1} needs $N_\text{MC}^\text{tr}$ Monte Carlo trials to stabilize for all choices of $\sigma$ and $\tau$. The accuracy of the data-driven closure does not improve for $N_\text{MC}^\text{tr}>4\times 10^4$; hence, $N_\text{MC}^\text{tr} = 4\times 10^4$ was used in the simulation results reported in Figs.~\ref{fig:coeff2}--\ref{fig:fhn_marg}. This number is approximately $20\%$ of the number of trials needed to compute the Monte Carlo solution $f_\text{MC}$ (i.e., $N_\text{MC} \approx 2 \times 10^5$).

\begin{figure}[htbp] 
\centering  
   \includegraphics[width=\linewidth]{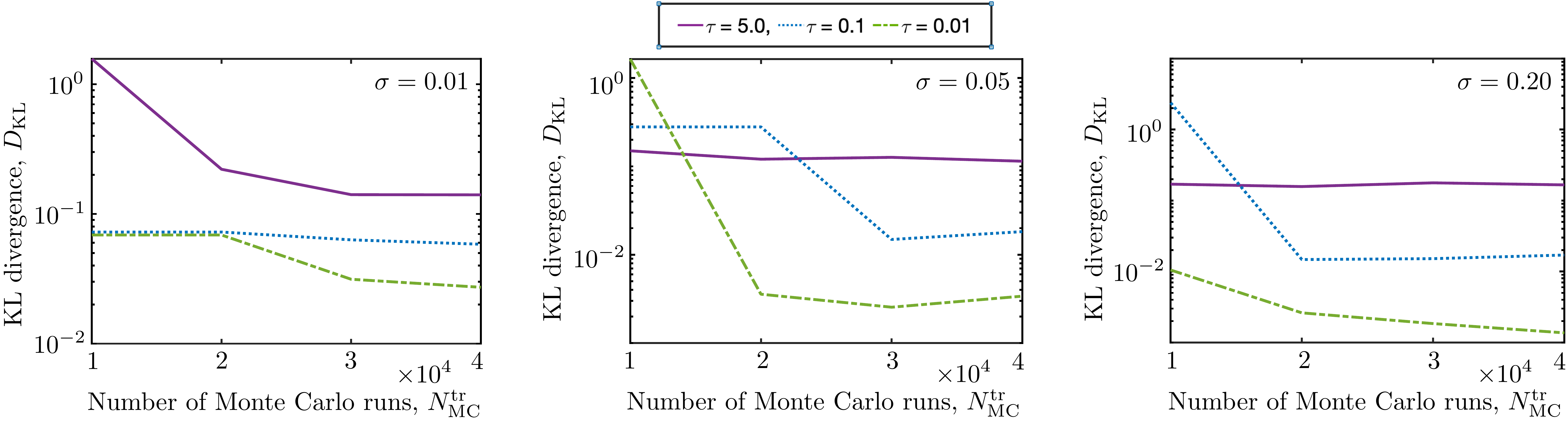}
\caption{Dependence of the KL divergence $D_\text{KL}(f_\text{MC} \| f_x)$, averaged over all temporal grid nodes, on the number of Monte Carlo runs used in the sparse regression, $N_\text{MC}^\text{tr}$.}
\label{fig:mc2_err}
\end{figure}

\subsubsection{Mutual Information}

Information transmission from a spiking neuron occurs via spike trains, i.e., a sequence of action
potentials. Observing a spike train can be thought of as gaining information  about the
sensory stimulus. Neuroscience regularly employs  information theory to formalize this notion on
an absolute scale so that quantifiable statements can be made about information transmission rates of
neurons~\cite{spikes}. Mutual information (MI) between two continuous random variables $x_i$ and $x_j$ is given in terms of the KL divergence of their joint PDF, $f_{x_i x_j}(X_i,X_j)$, and the product of their marginals, $f_{x_i}(X_i)$ and $f_{x_j}(X_j)$:
\[ I(x_i;x_j) \triangleq \int\int f_{x_ix_j}(X_i,X_j;t) \ln\left(\frac{f_{x_ix_j}(X_i,X_j;t)}{f_{x_i}(X_i;t)f_{x_j}(X_j;t)}\right) \text dX_i\text dX_j = D_\text{KL}(f_{x_i x_j} || f_{x_i}\otimes f_{x_j}) ; \] 
it is symmetric and nonnegative \cite{Mackay}. While either of the semi-local and data-driven PDF solutions from Sections~\ref{sec:semiloc2} and~\ref{sec:dd2} can be used to calculate MI between the membrane potential $x_1$ and the recovery variable $x_2$ in the FHN neuron, the dynamical system~\eqref{fhn1} must be modified in order to calculate MI between $x_1$ and the 
noise source $\xi$. Letting $ \zeta(t) = \s \cos\left(\sqrt{\frac{2}{\tau}}W_t\right)$, it follows via It\^o's formula~\cite[Theorem 4.1.2]{ksendal_2003} that $\xi$ and $\zeta$ satisfy the It\^o SDEs 
\begin{eqnarray} \label{sw_sde}
\text d\xi_t &=& -\frac{1}{\tau}\xi_t\, \text dt + \sqrt{\frac{2}{\tau}}\zeta_t \text d W_t,  \qquad
\xi(0) = 0,\notag\\
\text d\zeta_t &=& -\frac{1}{\tau}\zeta_t\, \text dt - \sqrt{\frac{2}{\tau}}\xi_t \text d W_t,  \qquad
\zeta(0) = 0.
\end{eqnarray}
Adding these two SDEs to~\eqref{fhn1} yields an exact four-dimensional Fokker-Planck equation. For any fixed time $t$, its solution is a joint PDF $f_{x_1 x_2 \xi \zeta}(X_1,X_2,\Xi,Z ;t)$ for the random processes $x_1(t)$, $x_2(t)$, $\xi(t)$, and $\zeta(t)$. Then, the instantaneous (point-wise in time) MI between the noise source and the membrane potential is found by calculating (via numerical quadrature) the joint PDF $f_{x_1 \xi}(X_1,\Xi;t)$ and the marginal PDFs $f_{x_1}(X_1;t)$ and $f_{\xi}(\Xi;t)$. This process involves solving a high-dimensional PDE and then numerically integrating out phase variables, which can be computationally expensive. In the spirit of the data-driven PDF method, a better approach is to directly learn reduced-order equations for the PDF(s) of interest (e.g., $f_{x_1 \xi}(X_1,\Xi;t)$) from data. Such an approach is presented in~\cite{BRENNAN2018281} for ODEs with random initial conditions, but can easily be generalized for SDEs driven by standard Wiener processes.

\begin{figure}[h!] 
\centering  
   \includegraphics[width=.4\linewidth]{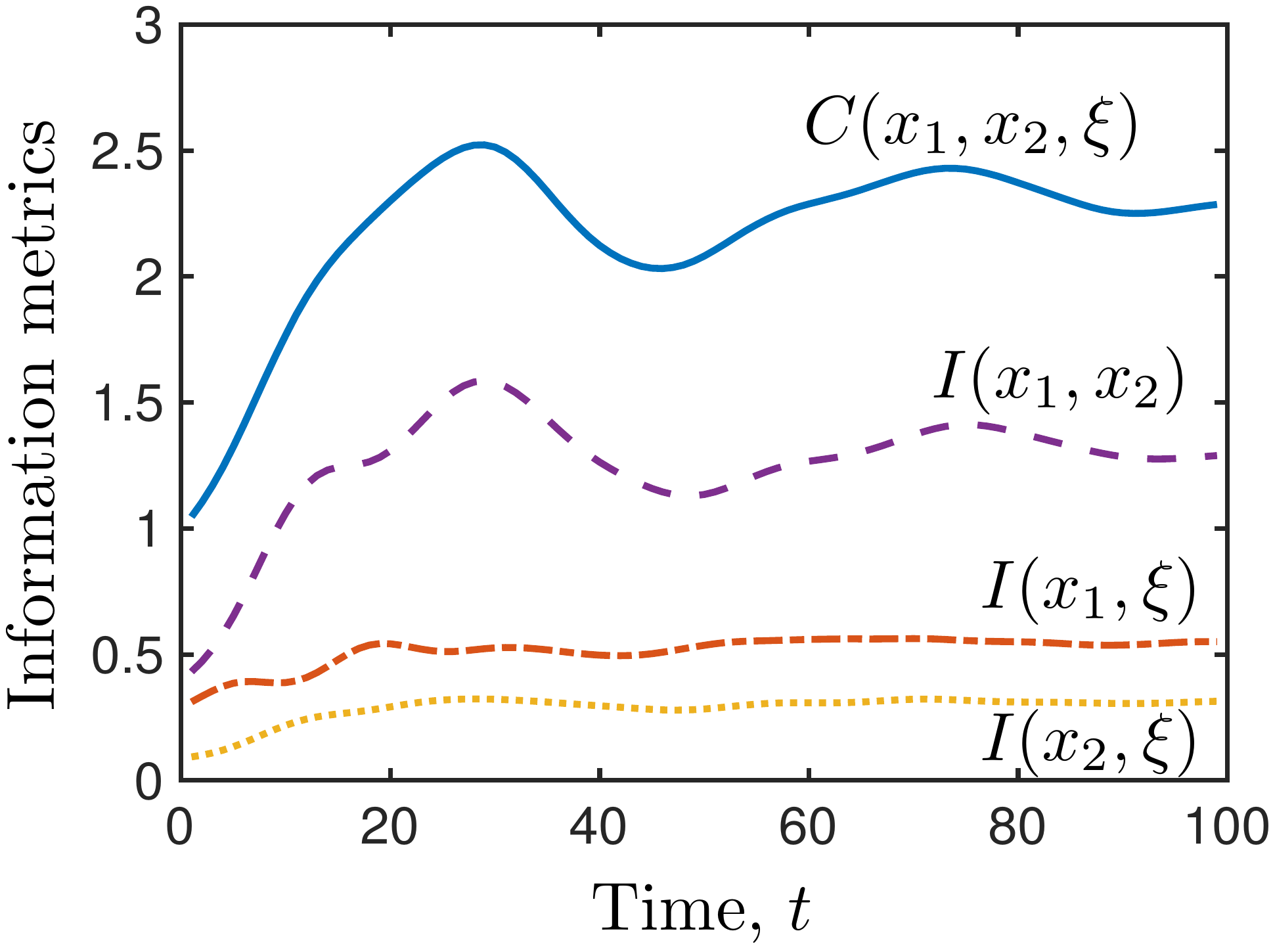}
\caption{Temporal evolution of MI, $I$, between the different FHN neuron states and of the total correlation, $C$, between all three states. The PDF equation~\eqref{pdfeq1} is closed with the data-driven closure, and the parameter values are set to $\sigma = 0.05$ and $\tau = 0.01$.}
\label{fig:mut_info}
\end{figure}

Fig.~\ref{fig:mut_info} displays the temporal evolution of MI between the various FHN states (including noise) for $\sigma = 0.05$ and $\tau = 0.1$. The MI initially increases in all cases: when the system begins to evolve, a portion of the joint PDF enters into a pseudo period due to the probability of a spike, and more information is gained about the coupling of states. However, there is a stark contrast in the evolution of $I(x_1;x_2)$ when compared to  $I(x_1;\xi)$ and  $I(x_2;\xi)$. The MI $I(x_1;x_2)$ oscillates according to the pseudo-periodic behavior of the joint PDF (Fig.~\ref{fig:fhn_plots}), i.e., the neuron's spikes. The MI between the membrane potential $x_1(t)$ and the recovery variable (inhibitory response) $x_2(t)$ is gained as the neuron spikes, but decreases as the neuron returns to the vicinity of its resting state, as one would expect~\cite{spikes}. Unlike $I(x_1;x_2)$,  both $I(x_1;\xi)$ and $I(x_2;\xi)$ are nearly constant in time after a slight initial increase. Hence, apart from the first spike, observing the membrane potential/recovery variable for a fixed time does not provide any more or less information about the noise source than any of the other temporal observations. The converse is also true since MI is symmetric. This behavior is attributed to the fact that the SDE for $\xi(t)$ is independent from the states $x_1(t)$ and $x_2(t)$. The noise $\xi$ can change the dynamics of $x_1$ and $x_2$ by perturbing the neuron from rest and causing it to fire; however, $x_1$ and $x_2$ have no effect on how the noise is generated. 

While MI is limited to measuring information transfer between two states/variables, its many generalizations allow one to do the same for $N$-many variables $x_1, \hdots, x_N$. One such generalization is total correlation \cite{totcor, Watanbe}. Like MI, it is defined in terms of the KL divergence of the variables' joint PDF, $f_{x_1\dots x_n}$, and the product of their marginals. Their total correlation is defined as
\[C(x_1, \hdots, x_N) = D_\text{KL}(f_{x_1 \hdots x_N} || f_{x_1} \otimes \hdots \otimes f_{x_N}). \] 
 Fig.~\ref{fig:mut_info} shows the total correlation between $x_1$, $x_2$, and $\xi$ for the FHN neuron as an example of another information-theoretic quantity that can be calculated via the PDF method.

\begin{remark}
The PDF method in Section 2 can be generalized in order to derive an equation for a joint PDF of system states at multiple times. This enables the calculation of information transfer rates between states at different times, as opposed to the instantaneous information transmission. For example, instead of calculating total correlation between the FHN states $x_1$ and $x_2$ and the noise source $\xi$ at any time $t$, one can compute the total correlation between these states at any two times $t_1$ and $t_2$. This would lead to the doubling of the number of dimensions in the PDF equation from three to six, and require deployment of numerical strategies for solutions of high-dimensional PDEs such as tensor-train methods~\cite{boelens-2018-parallel, boelens-2020-tensor}.
\end{remark}

\begin{remark}
Apart from computing information-theoretic metrics, the PDF method can be used to calculate PDFs for first exit/spike times and interspike-intervals. For example, the FHN neuron spikes (almost surely) when its path crosses the line $X_1 = 0$ \cite{yamakou19}. Defining the first exit/spike time to be $\tau_1 = \inf\{s>0: x_1(s) = 0\}$, its distribution can be calculated in terms of the cumulative distribution function (CDF) $F_\x(X_1,X_2;t)$.
When studying first spiking times, it is common to consider deterministic initial conditions in~\eqref{ivp2}. This setting corresponds to the initial condition for the PDF equation~\eqref{pdfeq1} being the Dirac delta function, which introduces significant numerical challenges~\cite{rutjens-2020-method}. This can be circumvented by deploying the CDF method (see \cite{Tartakovsky-2015-Method} and the references therein). It would result in a PDE for the CDF of system states, for which the deterministic initial condition in~\eqref{ivp2} translates into the Heaviside function as the initial CDF.
\end{remark}
 \section{Conclusions} 
 \label{sec:conclusions}
 Nonlinear dynamical systems provide a rich framework for understanding complex neuronal processes. Their stochastic analogues allow one to study the effects of externally imposed and/or internally generated noise on neuronal dynamics. They can also be used to quantify parametric uncertainty in underlying models. Noise sources in ODE-based neuron models are often treated as Gaussian white noise due to its convenient mathematical properties, one being that the joint PDF of system states satisfies exactly the Fokker-Planck equation. However, Gaussian white noise is not a biologically plausible noise source for many neuronal systems, for which colored noise characterized by a short correlation length is a more realistic representation.
 
Monte Carlo methods are often employed to calculate PDF solutions of stochastic and random ODEs, as well as information-theoretic quantities and distributions of various spiking statistics. They are easily implementable, yet computationally demanding. Efficient methods for general Langevin-type systems with colored noise have been studied extensively; however, most are not appropriate for neuron models. Methods utilizing only the first few moments of system states typically underperform because low-order moments can spend considerable time in a low probability state, illustrated in Fig. \ref{fig:fhn_plots} by the ensemble mean. Additionally, many methods that do characterize a full PDF solution, such as polynomial chaos expansions and stochastic finite elements, suffer from the curse of dimensionality when applied to systems with short correlation lengths. The methods of distributions, comprised of PDF and CDF methods, is the exception; this study utilizes the former to discover Fokker-Planck-type equations for stochastic neuron models. Such equations describe PDF dynamics of neuronal states, and in general, the states of general Langevin-type systems. While the PDF equations resulting from the method of distributions are exact for ODEs driven by white noise (i.e., SDEs), they are generally unclosed and require approximations in the presence of colored noise. We proposed two approaches for constructing such closures: nonlocal large eddy diffusivity (LED) closures and a data-driven closure relying on statistical inference; we apply each to study the stochastic NS-LIF and FHN neurons. 
 
Fully nonlocal LED closures result in derived integro-partial differential PDF equations that are difficult to discretize numerically and computationally demanding to solve unless the underlying RODEs exhibit weak nonlinearities and are low-dimensional, e.g., the NS-LIF model. For more complicated models, e.g., the FHN neuron, localization approximations are needed to render the PDF equation numerically computable. Full localization yields ``easy-to-solve'' linear partial-differential PDF equations, but introduces significant approximation error unless the noise intensity and correlation length are  small. Partial (semi-) localization provides an attractive trade-off between accuracy and computational complexity; however, this is likely to be problem dependent. Regardless, LED closures reduce computational costs by several orders of magnitude when compared to the standard Monte Carlo approach, with the local and semi-local LED closures being more efficient than the nonlocal one.

Unlike LED closures, the proposed data-driven closure needs only a few Monte Carlo trials and kernel density estimation to learn terms in the PDF equation from a dictionary of possible differentiable operators. The relevant derivative features and their coefficients are found by formulating an optimization problem via sparse regression, i.e., Lasso. If the number of equations in a neuron model is greater than one, the dimensionality of the regression problem is inherently high when considering arbitrary variable coefficients. It can be significantly reduced by approximating the coefficients with a polynomial basis expansion, which is sufficiently accurate for the models considered in this study. While more computationally expensive than the local and semi-local LED closures, the data-driven closure is easier to implement and still less expensive than the standard Monte Carlo approach. Its accuracy is comparable to the semi-local LED closure.

The method of distributions is a computationally efficient approach for finding PDF solutions to stochastic neuron models driven by colored noise. Post-processing of these solutions yields distributions of spiking statistics or information-theoretic quantities such as mutual information. For simple models, e.g., the NS-LIF neuron, all of the proposed closures yield consistent results and remain accurate over a wide range of noise intensities and correlation lengths. For models with more complicated dynamics, e.g., the FHN model and other two- or three-dimensional models such as Morris-Lecar, the semi-local LED and data-driven closures are likely to be in agreement and maintain good accuracy as long as $\sigma\tau$, the product of the noise intensity and correlation length, is not very large relative to the system's scale. For high-dimensional models, such as Hodgkin-Huxley (HH), the numerical discretization of PDF equations resulting from LED closures may be difficult, the data-driven closure is likely to give better results. 

 In a follow-up study, we will investigate both the HH neuron as well systems of coupled neurons, all driven by colored noise. In the latter case, we aim to use the method of distributions to determine how model complexity (i.e., using LIF versus FHN versus HH neurons, etc.) affects information content and transmission in networks of neurons.  Such a study will require deployment of numerical strategies for solutions of high-dimensional PDEs such as tensor-train methods~\cite{boelens-2018-parallel, boelens-2020-tensor, Dektor_2021, rodgers2021adaptive}.

\section*{Acknowledgements}

This work was supported in part by Air Force Office of Scientific Research under award number  FA9550-18-1-0474, National Science Foundation Graduate Research Fellowship under Grant No. DGE 1752814, and by UC Berkeley SURF L\&S Program. A special thanks to Michael R. DeWeese and his useful comments regarding the manuscript.


\bibliography{pdfmethod}

\appendix

\section{Derivation of PDF Equations}
\label{regularize}

\subsection{Regularization Argument}

Let us define $\Pi_\e(\mathbf X,t)$, a regularized version of $\Pi(\mathbf X,t)$ in~\eqref{rawpdf}, as 
\begin{equation}\label{eq:Pi}
\Pi_\e(\X,t) \triangleq (\eta_\e \star \Pi)(\X,t) \triangleq \int_{\Rn} \eta_\e(\X-\Y)\Pi(\Y,t) \text d\Y = \eta_\e(\X-\x),
\end{equation}
where the last equality holds by the definition of $\Pi(\Y,t)$ and the sifting property of the Dirac distribution.
The standard positive mollifier $\eta_\e \in \mathscr{C}_0^{\infty}(\Rn)$ satisfies the conditions of symmetry, $\eta_\e(\X-\x)= \eta_\e(\x-\X)$, and scaling
\begin{equation}
\eta_\e(\Y) \triangleq \frac{\e^{-N}}{\int\eta \text d\Y}\,\eta\left(\frac{\Y}{\e}\right), \quad \text{where } \eta(\Y) 
\triangleq \begin{cases} \exp\left(\frac{1}{|\Y|^2-1}\right) &\text{ if } |\Y| < 1\\ 0 &\text{ if } |\Y| \ge 1.\end{cases}
\end{equation}
Following the arguments from~\cite{evans}, one can show that $\Pi_\e$ is a smooth approximation of $\Pi$. 
Let $\phi(\X,t) \in \mathscr{C}_c^1(\Rn\times [0,\infty))$. It follows from~\eqref{eq:Pi} that
\begin{align}\label{I1}
\mathcal{I} \triangleq \int_0^{\infty}\int_\Rn \Pi_\e(\X,t)\frac{\partial \phi}{\partial t}(\X,t) \text d\X \text dt =  \int_0^{\infty}\int_\Rn \eta_\e(\X-\x)\frac{\partial \phi}{\partial t}(\X,t) \text d\X \text dt.
\end{align}
Integrating by parts in $t$ and applying the sifting property gives 
\begin{align*}
\mathcal{I}  = \;& \int_0^{\infty}\int_\Rn \dot{\eta}_\e(\X-\x)\v(\x,t,\w)\phi(\X,t) \text d\X \text dt - \int_\Rn  \eta_\e(\X-\x_0) \phi(\X,0) \text d\X,\\
= \;& \int_0^{\infty}\int_\Rn \int_\Rn \dot{\eta}_\e(\X-\Y)\v(\Y,t,\w)\Pi(\Y,t)\phi(\X,t) \text d\Y \text d\X \text dt - \int_\Rn  \Pi_\e(\X,0)\phi(\X,0) \text d\X,
\end{align*}
where $\dot{\eta}_\e( \cdot)$ is the derivative of $\eta_\e(\cdot)$. According to the Gauss-Ostrogradsky theorem in $\X$,
\begin{align}\label{I2}
\mathcal{I}  =  -\int_0^{\infty}\int_\Rn (\eta_\e \star \v\Pi)(\X,t) \grad_\X \phi(\X,t) \text d\X \text dt - \int_\Rn  \Pi_\e(\X,0)\phi(\X,0) \text d\X.
\end{align}
It follows from~\eqref{I2} and~\eqref{I1} that, for any $\phi \in \mathscr{C}_c^1(\Rn\times [0,\infty))$, 
\begin{align*}
\int_0^{\infty}\int_\Rn \Pi_\e\frac{\partial \phi}{\partial t} \text d\X \text dt + \int_0^{\infty}\int_\Rn (\eta_\e \star \v\Pi) \grad_\X \phi \text d\X \text dt + \int_\Rn  \Pi_\e(\X,0)\phi(\X,0) \text d\X = 0.
\end{align*}
By standard arguments, taking the limit $\e \rightarrow 0$ gives
\begin{align} \label{weak_sol}
\int_0^{\infty}\int_\Rn \Pi\frac{\partial \phi}{\partial t} \text d\X \text dt + \int_0^{\infty}\int_\Rn (\v\Pi) \grad_\X \phi \text d\X \text dt + \int_\Rn  \Pi(\X,0)\phi(\X,0) \text d\X \text dt = 0,
\end{align}
 Hence, $\Pi$ is the distributional solution to (\ref{clpi}). 

 
\subsection{LED closure derivation}\label{closuresection}
Let $L$ denote the linear operator 
\[
L\Pi \triangleq  \frac{\d\Pi}{\d t} + \grad_{\X} \cdot (\langle\v\rangle\Pi) 
\]
together with the initial and boundary conditions 
\begin{align} 
\Pi(\X,0) = \delta ( \mathbf x^0 - \mathbf X) \quad\mbox{ and } \quad \lim_{|\X|\to \infty} \Pi(\X,t) = 0. \label{ibcpi}
\end{align}
Let $\hat L$ denote the adjoint of $L$, i.e.,
\[
\hat L\Psi \triangleq  -\frac{\d\Psi}{\d s} -\langle\v\rangle\ \cdot   \grad_{\Y} \Psi
\]
with vanishing boundary condition at infinity and vanishing terminal condition $\Psi (\Y,t) = 0$. Finally, let $G \equiv G(\X,t; \Y,s)$ be the Green's function of $\hat L$; it is defined as a solution of  
\begin{align}
\hat LG = \delta(\X-\Y)\delta(t-s), \label{Geq}
\end{align}
with  vanishing boundary condition at infinity and terminal condition $G(\X,t;\Y,t) = 0$. We use the method of characteristics to compute this Green's function. The characteristics of~\eqref{Geq}, $ \boldsymbol\varphi \equiv  \boldsymbol\varphi(r)$ with $s \le r \le t$, satisfy 
 \begin{align}
 \frac{\text d\boldsymbol\varphi}{\text dr} = \langle\v(\boldsymbol\varphi,r)\rangle, \qquad r \in (s,t), \label{cceq}
 \end{align}
 subject to the initial condition
 \begin{align}
 \boldsymbol\varphi(s ) = Y. \label{ccini}
 \end{align}
 It induces the associated flow $\boldsymbol\Phi$ as $\boldsymbol\varphi(r) \equiv \boldsymbol\Phi(r;\Y,s)$.
 Along these characteristics,~\eqref{Geq} reduces to 
 \[
 \frac{\text d}{\text dr}G(\X,t ; \boldsymbol\varphi, r) = -\delta(t - r)\delta(\X - \boldsymbol\varphi), \qquad  r \in (s,t);
 \]
 with  $G(\X,t ; \boldsymbol\Phi(t ; \Y,s),t) = 0$. Its solution gives the Green's function,
 \begin{align}
 G(\X,t ; \Y,s) = \H(t - s)\,\delta(\X - \boldsymbol\Phi(t ; \Y,s)), \label{greenf}
 \end{align}
 where $\H$ is the Heaviside function.  
 
Elementary manipulations lead to the equation for $\Pi'$,
\begin{align}
L\Pi' = -\grad_{\X} \cdot (\v'\Pi - \langle\v'\Pi'\rangle), \label{piprime}
\end{align}
with homogeneous initial conditions and vanishing boundary conditions. Rewriting (\ref{piprime}) in terms of $\Y$ and $s$, multiplying by $G(\X,t ; \Y,s)$, and integrating by parts, we obtain
\[
\int_{0}^{t}\!\!\int_{\Rn} GL\Pi' \text d\Y \text ds = -\int_{\Rn}  G(\X,t;\Y,0) \Pi'(\Y,0) \text d\Y +  \int_{0}^{t} \int_{\Rn} \Pi'\hat{L}G \text d\Y \text ds,  
 \]
where the additional boundary terms cancel because of  the terminal condition on $G$ and the vanishing boundary condition (\ref{ibcpi}) at infinity. Accounting for~\eqref{Geq} and~\eqref{piprime}, 
\begin{align*}
\begin{split}
\Pi'(\X,t) &=  \int_{\Rn}  G(\X,t;\Y,0) \Pi'(\Y,0) \text d\Y \\
&+ \int_{0}^{t}\!\! \int_{\Rn} G(\X,t ; \Y,s) \grad_{\Y} \cdot \biggl( \langle\v'(\Y,s)\Pi'(\Y,s) \rangle -\v'(\Y,s)\Pi(\Y,s) \biggr) \text d\Y \text ds.
 \end{split}
\end{align*} 
The first integral vanishes if the initial condition on $\mathbf x^0$ is deterministic. An exact, albeit unclosed, equation for the stochastic flux $\langle\v'\Pi'\rangle$  is obtained by multiplying the previous relation with $\v'(\X,t)$ and taking the ensemble mean,
\[
\begin{split}
\langle\v'\Pi'\rangle(\X,t) &= \int_{\Rn}  G(\X,t;\Y,0) \langle \v'(\X,t) \Pi'(\Y,0)\rangle \text d\Y  \\
& -\int_{0}^{t}\!\! \int_{\Rn} G(\X,t ; \Y,s) \grad_{\Y} \cdot \langle\v'(\X,t)\v'^\top(\Y,s)\Pi(\Y,s) \rangle \text d\Y \text ds.
\end{split}
\]
The LED  closure is constructed by setting the first integral to zero, resulting in a second-order (in $\s\tau$) computable relationship  
\begin{align}
\langle\v'\Pi'\rangle(\X,t) \approx - \int_{0}^{t} \!\!\int_{\Rn} G(\X,t ; \Y,s) \grad_{\Y} \cdot \Big(\langle\v'(\X,t)\v'^\top(\Y,s) \rangle  f_\y(\Y;s)\Big)\text d\Y \text ds. \label{fluxeq}
 \end{align} 
 Substituting $G$ from~\eqref{greenf} and setting 
  \[
 q(\X,\Y,t,s) \triangleq   \grad_{\Y} \cdot \Big(\langle\v'(\X,t)\v'^\top(\Y,s) \rangle  f_\y(\Y,s)\Big),
  \]
  we obtain
  \begin{align*}
\langle\v'\Pi'\rangle(\X,t) \approx & - \int_{0}^{t} \!\!\int_{\Rn}  \delta(\X-\boldsymbol\Phi(t;\Y,s)) \, q(\X,\Y, t,s) \text d\Y \text ds \\
\approx & - \int_{0}^{t} \!\! \int_{\Rn} \delta(\X-\boldsymbol\varphi) \, q(\X,\boldsymbol\Phi(s;\boldsymbol\varphi, t),t,s)\,\left|\frac{\partial \boldsymbol\Phi(s;\boldsymbol\varphi,t)}{\partial \boldsymbol\varphi}\right| \text d\boldsymbol\varphi \text ds.
  \end{align*}
 The Jacobian determinant in this expression is made explicit by invoking the Liouville-Ostrogradsky lemma. It is reproduced and proved below because we were not able to find an appropriate reference. 
 
 \begin{lem}[Liouville-Ostrogradsky]
 \label{lo}
 Consider an initial value problem 
 \begin{align}\label{lo1}
 \frac{\text d\boldsymbol\chi}{\text ds} =\mathbf g(\boldsymbol\chi, s),  \qquad
 \boldsymbol\chi(t) = \boldsymbol\xi, 
 \end{align}
 where $\mathbf g:\mathbb R^N\times \mathbb R \to \mathbb R^N$ is a given deterministic function, and $\boldsymbol\xi$ is a possibly random initial condition.   Both $\mathbf g$ and $\boldsymbol\xi$ are such that, for any realization of $\boldsymbol\xi$,   the above system admits a unique solution for the range of time under consideration. Then, if $\boldsymbol\Phi$ is the corresponding flow, i.e., $\boldsymbol\chi(s) \equiv \boldsymbol\Phi(s; \boldsymbol\xi, t)$, we have 
\[
 \left| \frac{\d\boldsymbol\Phi(s ; \boldsymbol\xi,t)}{\d\boldsymbol\xi} \right| = \exp \left(- \int_s^t \grad_{\boldsymbol\chi} \cdot \mathbf g(\boldsymbol\chi(r),r) \text dr \right).
 \]
 \end{lem}
 \begin{pf} 
Following the arguments of Section~\ref{sec:pdf}, we show that $f$, the PDF of $\boldsymbol \chi$, satisfies
 \[
 \frac{\partial f}{\partial t} + \nabla_{\boldsymbol\chi}  \cdot (\mathbf g f) = 0. 
 \]
By construction, the restriction of $f$ on characteristic curves, $\mathcal P(s) = f(\boldsymbol\chi(s);s)$, satisfies
 \[
 \frac{\text d\mathcal P}{\text ds} = -\mathcal P \,\nabla_{\boldsymbol\chi} \cdot \mathbf g.
 \] 
Hence, 
 \[
 \mathcal P(s) = \mathcal P(t) \exp\left( - \int_t^s \nabla_{\boldsymbol\chi} \cdot \mathbf g(\boldsymbol\chi(r),r) \text dr\right). 
 \]
 On the other hand, treating the flow $\boldsymbol\Phi$ as a one-to-one change of variables,  
 \[
 \mathcal P(s) = \left|\frac{\partial \boldsymbol\Phi^{-1} }{\partial \boldsymbol\chi} \right|\mathcal P(t).
 \]
 Combining these two expressions, and taking the inverse completes the proof.
 \qed
 \end{pf}
  
We apply the Liouville-Ostrogradsky lemma to $\boldsymbol\chi$ and $\boldsymbol\Phi$ defined, respectively, as the solution of and the flow corresponding to~\eqref{lo1} with  $\mathbf g(\chi,s) =  \langle\v(\boldsymbol\chi,s)\rangle$ and $\boldsymbol\xi = \X$. This yields 
\begin{subequations}\label{nonlocalclosure}
   \begin{equation}
\langle\v'\Pi'\rangle(\X,t) \approx - \int_{0}^{t}   \J(s ; \X,t)  \grad_{\boldsymbol\Phi} \cdot \Big(\langle\v'(\X,t)\v'^\top(\boldsymbol\Phi(s;\X, t),s) \rangle  f_\x(\boldsymbol\Phi(s;\X, t);s)\Big)   \text ds, 
\end{equation}
where 
\begin{align}
 \J(s ; \X,t) =  \exp \left(- \int_s^t \grad_{\boldsymbol\chi} \cdot \langle\mathbf v(\boldsymbol\chi(r),r)\rangle \text dr \right).
 \end{align}
 \end{subequations}
Substituting~\eqref{nonlocalclosure} into~\eqref{pdfeq1} gives an integro-differential equation for $f_\x$. Its numerical solution poses significant challenges as the resulting problem is local neither in time nor in space. 

The localization inherent in the classical LED theory \cite{kraichnan-1987-eddy} assumes that $f_\x$ 
and its spatial derivatives are approximately constant over the correlation-length interval 
$(t-\tau, t)$. This gives rise to approximations 
$f_\x(\boldsymbol{\Phi}(s; \mathbf X, t); s) \approx f_\x(\mathbf X; t)$ and 
$\nabla_{\boldsymbol{\Phi}} f_\x(\boldsymbol{\Phi}(s; \mathbf X, t); s) \approx
\nabla_{\mathbf X} f_\x(\mathbf X; t)$, so that
\begin{equation}
\langle\v'\Pi'\rangle(\X,t) \approx \boldsymbol{\mathscr{V}}(\X,t)f_\x(\X;t) - \mathscr{D}(\X,t)\grad_{\X}\,f_\x(\X;t), \label{led}   
\end{equation}
where $\mathscr{V}$ and $\mathscr{D}$ are the LED drift velocity vector and diffusion tensor
\begin{align}
\mathscr{V}(\X,t) \triangleq& -  \int_{0}^{t} \J(s ; \X,t)\langle\v'(\X,t)\grad_{\boldsymbol\Phi} \cdot \v'^\top(\boldsymbol\Phi(s; \X,t),s)\rangle \text ds, \label{leddrift} \\
\mathscr{D}(\X,t) \triangleq&  \int_{0}^{t} \J(s ; \X,t) \langle\v'(\X,t)\v'^\top(\boldsymbol\Phi(s; \X,t),s) \rangle \text ds. \label{leddiffusion}
\end{align} 
Substituting~\eqref{led}--\eqref{leddiffusion} into (\ref{pdfeq1}) results in the classical LED equation for the joint PDF $f_\x$:
\begin{eqnarray}
\frac{\d f_\x}{\d t} + \grad_{\X} \cdot \Bigl((\langle\v\rangle + \V)f_\x\Bigr) = \grad_{\X} \cdot (\D\,\grad_{\X}\,f_\x). \label{ledclassic}
\end{eqnarray} 
The shortcomings of the above LED closure---most importantly, its limited validity to short correlation lengths---are investigated in~\cite{bst,Maltba,ye-2004-nonlocal}, among many other studies.

\subsection{Semi-local LED closure}
\label{app:semilocal}
 
This closure is based on the mean-field advection problem
\begin{align}
\frac{\d f_\x}{\d s} + \nabla_{\X} \cdot (\left<\v\right>f_\x) = 0, \qquad  s<t \label{meanf}
\end{align}
and is equivalent to
\begin{align}
f_\x(\boldsymbol\Phi(s ; \X,t);s) = \J^{-1}(s;\X,t)f_\x(\X;t). \label{meanft}
\end{align}
Equation (\ref{meanft}) gives a local approximation for $f_\x$; substituting it into~\eqref{nonlocalclosure} yields the semi-local closure
\begin{align}
\langle\v'\Pi'\rangle(\X,t) \approx -  \int_{0}^{t} \J(s ; \X,t) \grad_{\boldsymbol\Phi} \cdot\Big( \langle\v'(\X,t)\v'^\top(\boldsymbol\Phi(s ;\X,t),s) \rangle   \J^{-1}(s;\X,t)f_\x(\X;t) \Big) \text ds.   \label{ledmod}
\end{align}
The PDF $f_\x$ in (\ref{ledmod}) has been localized from $(\boldsymbol\Phi(s;\X,t);s)$ to $(\X;t)$, but the operator $\nabla_{\boldsymbol\Phi}\cdot$ has not. We now approximate $\nabla_{\boldsymbol\Phi}$ in terms of $\nabla_{\X}$,
 \begin{align}
 \nabla_{\X} =\boldsymbol\Psi^\top \nabla_{\boldsymbol\Phi} \quad\text{with}\quad
 \boldsymbol\Psi_{ij} \triangleq \frac{\partial \boldsymbol\Phi_i}{\partial \X_j} (s; \X,t).
 \label{nablaeq}
 \end{align}
Elementary calculus shows that the transpose of the sensitivity matrix of the flow, $\boldsymbol\Psi(s; \X,t) $, satisfies
 \begin{align*}
 \frac{\text d \boldsymbol\Psi}{\text ds} = \mathbf J\boldsymbol\Psi, \qquad
 \boldsymbol\Psi(t; \X,t) = \mathbb I,
 \end{align*}
where $\mathbf J(\mathbf Y,s) \triangleq  \partial \langle \mathbf v(\mathbf Y,s)\rangle/\partial \mathbf Y$ is the Jacobian of the mean-field velocity, and $\mathbb I$ is the $N\times N$ identity matrix. Consequently, $\boldsymbol\Psi$ can be expressed as
 \begin{align*}
 \boldsymbol\Psi(s; \X,t) = \;& \text{OE}[\mathbf J](s) \\
 \triangleq \;& \mathbb I + \int_t^s\! \mathbf J( \boldsymbol\Phi(s_1;\X,t),s_1) \text ds_1 + \int_t^s\!\!\int_t^{s_1} \mathbf J( \boldsymbol\Phi(s_1;\X,t),s_1)  \mathbf J( \boldsymbol\Phi(s_2;\X,t),s_2) \text ds_1 \text ds_2 +\dots,
 \end{align*}
 where $\text{OE}$ is the ordered exponential. Using the approximation $\mathbf J( \boldsymbol\Phi(s;\X,t),s) \approx \mathbf J(\X,t)$, the ordered exponential simplifies to the matrix exponential. Hence, we obtain
 \[
 \boldsymbol\Psi \approx \exp((s-t)\mathbf J(\X,t)),
 \]
 and, accounting for~\eqref{nablaeq},
\begin{equation} \label{nabla_app}
\nabla_{\boldsymbol\Phi} \approx \exp((t-s)\j(\X,t))^\top \nabla_{\X}.
\end{equation}
Substituting (\ref{nabla_app}) into (\ref{ledmod}), we obtain the approximate semi-local closure~\eqref{ledmod1}.

\end{document}